\documentclass{article}
\usepackage[english]{babel}
\usepackage[a4paper]{geometry}
\usepackage{enumerate}
\usepackage{amsmath}
\usepackage{amsthm}
\usepackage{amssymb}
\usepackage{hyperref}
\usepackage{amssymb}
\usepackage{color}
\usepackage[utf8]{inputenc}
\usepackage{graphicx}
\usepackage{tikz}
\usepackage{tikz-cd}
\usepackage{verbatim} 
	\usepackage{comment}
	\usetikzlibrary{decorations.markings}
	\usepackage{bbm} 
	\usepackage{xfrac} 
	\usepackage{tensor} 
	\usepackage{listings}  
	\usepackage{matlab-prettifier} 
	
	\usepackage{pgfplots}              
	\pgfplotsset{compat=newest}
	\usepgfplotslibrary{fillbetween}

\newtheorem{theorem}{Theorem}[section]  

\newtheorem{prop}[theorem]{Proposition}
\newtheorem{lemma}[theorem]{Lemma}

\numberwithin{equation}{section}
\theoremstyle{definition}

\newtheorem{rem}[theorem]{Remark}
\newtheorem{as}{Assumption}

\newcommand{\ep}[1]{\mathrm{e}^{#1}}
\newcommand{\iu}{\mathrm{i}}
\newcommand{\esup}{\mathrm{ess\,sup}}
\newcommand{\einf}{\mathrm{ess\,inf}}
\newcommand{\MW}{\mathcal{M}_W}

\newcommand{\caB}{\mathcal{B}}
\newcommand{\caH}{\mathcal{H}}
\newcommand{\caQ}{\mathcal{Q}}
\newcommand{\caN}{\mathcal{N}}
\newcommand{\caE}{\mathcal{E}}
\newcommand{\caF}{\mathcal{F}}
\newcommand{\caG}{\mathcal{G}}
\newcommand{\caR}{\mathcal{R}}
\newcommand{\caO}{\mathcal{O}}

\newcommand{\be}{\beta^{(\varepsilon)}}
\newcommand{\de}{\delta^{(\varepsilon)}}
\newcommand{\ze}{\zeta^{(\varepsilon)}}
\newcommand{\ps}{\psi^{(\varepsilon)}}
\newcommand{\ph}{\phi^{(\varepsilon)}}
\newcommand{\thee}{\theta^{(\varepsilon)}}
\newcommand{\chie}{\chi^{(\varepsilon)}}
\newcommand{\xie}{\xi^{(\varepsilon)}}
\newcommand{\et}{\eta^{(\varepsilon)}}
\newcommand{\Ham}{\mathcal{H}^{(\varepsilon)}}
\newcommand{\re}{r^{(\varepsilon)}}

\newcommand{\R}{\mathbb{R}}
\newcommand{\C}{\mathbb{C}}
\newcommand{\Rt}{\widetilde{\mathcal{R}}}
\newcommand{\Gt}{\widetilde{G}}

\title{Nonlinear adiabatic theorem for the Gross-Pitaevskii equation}

\begin{document}
		\author{Sven Bachmann$^1$, Severin Schraven$^2$ \\
		\\
		$^1$ Department of Mathematics, The University of British Columbia, \\
		Vancouver, BC V6T 1Z2, Canada \\
		$^2$ Department of Mathematics, Technical University of Munich, Garching, Germany}
	
	\maketitle
	
\begin{abstract}
We prove a nonlinear adiabatic theorem for the repulsive Gross-Pitaevskii (GP) equation with a time-dependent trapping potential $V(x,t)$. If either (i) $V(\cdot,t)$ is smooth, subquadratic in space and with uniformly bounded spatial derivatives of order $\geq 2$, or (ii) $V(\cdot,t)$ is a uniformly bounded perturbation of a fixed non-negative potential which is asymptotically $|x|^m, m\geq 2$, then the GP equation has a global solution, which remains uniformly close to the instantaneous minimizer of the time-independent GP functional, provided the initial condition is a ground state. In fact, we show that the existence of a solution of the GP equation is the real bottleneck: If existence is assumed, then adiabaticity holds under very minimal assumptions. 
\end{abstract}

\section{Introduction}

The adiabatic principle refers to a general fact that a dynamical system initially in a stationary state evolves under slowly varying conditions so that it remains at all times close to an instantaneous stationary state. In quantum mechanics, this principle was realized as a theorem first by Born and Fock~\cite{BornFock}, and in generality under a spectral gap condition by Kato~\cite{kato1950adiabatic}, see also~\cite{Avron2012} for a very general result applying to contracting evolutions in Banach space.

For the case of a quantum gas with $N$ particles, typical error bounds are indeed small as a function of how slowly the system's parameters evolves, but divergent in the large $N$ limit, thereby making the adiabatic principle fail if the thermodynamic limit is taken first. 

This issue was overcome in the setting of infinite quantum lattice systems in~\cite{bachmann2018adiabatic}, where the essential tools are propagation estimates (the Lieb-Robinson bound~\cite{NS}) and a local spectral flow~\cite{hastings2004locality, BMNS}. These hold for spin systems and for lattice fermions~\cite{MonacoTeufel}, but the Lieb-Robinson bound does not hold for bosons or continuum systems at the same level of generality, see~\cite{LemmRubiliani} for a review of recent results in that direction. As a consequence, there is still no general adiabatic theorem for thermodynamic systems in the continuum.

Bosons in the thermodynamic limit are known to exhibit Bose-Einstein condensation in their ground states. While this is a theorem only for non-interacting Bose gases, Bose-Einstein condensation can also be obtained in an appropriate scaling limit, where the Bogoliubov approximation gives rise to the Gross-Pitaevskii (GP) theory. The relevant quantity is the minimizer of the GP functional.
\begin{equation}\label{GP functional}
	\caE_{\mathrm{GP}}(u) = \int_{\mathbb{R}^3} \left(\vert \nabla u\vert^2 + V \vert u \vert^2 + \frac{\kappa}{2} \vert u\vert^4\right).
\end{equation} 
Here the scattering length $\kappa>0$ depends only on the microscopic interaction potential while $ V$ is a trapping potential in the sense that it diverges as $\vert x\vert\to\infty$. The minimizer has the physical interpretation of the one-particle orbital which has macroscopic occupation. More mathematically, the reduced $k$-particle density matrices of the many-body ground states converges to the $k$-fold tensor product of the projection onto the GP minimizer (see \cite{lieb2002proof} and \cite[Theorem 2.2]{erdHos2010derivation}). Importantly for the present work, the convergence of the reduced density matrix survives in a dynamical setting. This was first established when the trap is completely switched off~\cite{ESYNoTrap}, and subsequently~\cite{ErdosSchleinYau,Pickl2010} with a time-independent external potential $V(x)$: If the bosonic gas starts in a condensate, then it remains in a condensate made of solutions of the dynamical GP equation. 
\begin{equation} \label{GP time-dependent eq}
	i\partial_t \psi(x,t) = (-\Delta + V(x) + \kappa \vert \psi(x,t) \vert^2) \psi(x,t).
\end{equation}

With this in mind, it is natural to ask the adiabatic question for bosons in the thermodynamic limit directly at the effective level of the GP equation. For this, we consider a time-dependent external potential $V_t$ and define the corresponding time-dependent linear operator
\begin{equation*}
\widetilde{H}_t(\varphi) = -\Delta + V_t(x) + \kappa \vert \varphi(x)\vert^2.
\end{equation*}
The slow variation hypothesis is introduced through the usual scaling $s = \varepsilon t$ with $s\in[0,1]$ and $\varepsilon\ll1$. In the rescaled variable $s$, the GP equation reads
\begin{equation} \label{GP rescaled eq}
	\iu \varepsilon \frac{d}{ds}\psi^{(\varepsilon)}(x,s) = \widetilde{H}_s\big( \psi^{(\varepsilon)}(x,s)\big) \psi^{(\varepsilon)}(x,s).
\end{equation}
At every fixed $s$, the GP functional~\eqref{GP functional} has a minimizer, which is the solution of
\begin{equation}\label{GP Minimizer}
	\widetilde{H}_s(\omega_s(x)) \omega_s(x) = \mu_s \omega_s(x)
\end{equation}
where $\mu_s = \min \sigma(-\Delta+V_s+\kappa\vert \omega_s \vert^2)$.

In this setting, the adiabatic principle can be phrased as follows. Assuming for now that the dynamical GP equation~\eqref{GP rescaled eq} has a unique solution on $s=[0,1]$ with initial condition $\psi^{(\varepsilon)}(x,0) = \omega_0(x)$ and that the minimizer given in~\eqref{GP Minimizer} is unique, then 
\begin{equation*}
    \left\Vert \ep{\frac{\iu}{\varepsilon}\int_0^s\mu_\tau d\tau}\psi^{(\varepsilon)}(x,s) - \omega_s(x)\right\Vert \leq C\varepsilon,
\end{equation*}
where $\ep{\frac{\iu}{\varepsilon}\int_0^s\mu_\tau d\tau}$ is a quickly oscillating dynamical phase.

The existence of a solution of~\eqref{GP rescaled eq} is not known in general. We will prove it in two cases. One goes back to~\cite{carles2009nonlinear} and has a subquadratic potential $V$, see Theorem~\ref{thm:subquadratic} below. The second one, Theorem~\ref{thm:poly growth} is an adaptation of results in \cite{carles2015sharp}: the time-dependent potential is a bounded perturbation of a reference potential, which has polynomial growth at infinity.

The validity of the adiabatic theorem holds in much broader generality, if the dynamics is assumed to exist: This is the content of Theorem~\ref{thm:conditional}, where the only assumptions are trapping and sufficient smoothness in time of the potential. All of these results will be derived from a general, abstract setting that does not require the linear part to be explicit, in particular it needs not be a Schr\"odinger operator, in which the adiabatic theorem holds for the GP nonlinearity. This is the content of Theorem~\ref{thm:abstract}.

Finally, let us comment on closely related results~\cite{GangGrech} and \cite{fermanian2020nonlinear}. On the one hand, \cite{fermanian2020nonlinear} proves an adiabatic theorem for a class of bounded non-nonlinearities which depend only on a fixed finite dimensional projection of the wavefunction, and under a strong spectral assumption on the eigenvalue counting function for the linear part. The results apply to all spectral subspaces, not only the ground state manifold. However, the spectral condition excludes Schr\"odinger operators with a confining potential in $\R^3$ because of Weyl asymptotics, and the allowed non-linearity does not cover the physically relevant cubic one of the GP theory. On the other hand, \cite{GangGrech} consider the GP equation, but with a very weakly trapping potential in the sense that $-\Delta + V_s$ admits exactly one eigenvalue below the continuum, separated by a gap, for all $s$. Moreover, the initial condition is assumed to have small $L^2$ mass. The reason for the validity of the adiabatic approximation there is of a very different nature than here: the results of~\cite{GangGrech} rely on scattering and dispersive estimates which are unavailable in the trapped case. We also mention earlier work with weak nonlinearities \cite{carles2011nonlinear, hari2013coherent, sparber2016weakly}.

\section{Results}

We fix now a time-independent potential $W(x)$ which will serve as a reference and satisfies the following assumptions.

\begin{as} \label{as: confining potential}
    The potential $W\in L_\mathrm{loc}^\infty(\mathbb{R}^3)$ satisfies $W\geq 1$ and
    \begin{equation}
        \lim_{R\rightarrow \infty} \einf _{\vert x \vert \geq R} W(x)=\infty. 
    \end{equation}
\end{as}

We specify now which type of time-dependent perturbations we can allow in order to get a nonlinear adiabatic theorem.

For this we introduce a weighted $L^\infty$ space
\begin{equation}
    \mathcal{M}_\mathrm{W}=\left\{ f\in L^\infty_\mathrm{loc}(\mathbb{R}^3) \ : \ \Vert f \Vert_{\mathcal{M}_W} = \esup_{x\in \mathbb{R}^3} \frac{\vert f(x)\vert }{W(x)}<\infty\right\}.
\end{equation}
Furthermore, we will denote by $\mathcal{Q}_W$ the form domain of $-\Delta+W(x)$, i.e.
\begin{equation}
    \mathcal{Q}_W = \left\{ f\in H^1(\mathbb{R}^3) \ : \ \Vert f \Vert_{\mathcal{Q}_W}^2 = \int \left[ \vert \nabla f(x)\vert^2+  W(x) \vert f(x) \vert^2 \right] dx < \infty \right\}.
\end{equation}
We want two types of conditions: firstly that the form domain of our Schrödinger operator is time-independent and that the change of the external potential is sufficiently regular. 

\begin{as} \label{as: form}
\begin{enumerate}
    \item There exists $c,C,M>0$ such that 
\begin{equation} \label{eq: form1}
        c W \leq V(\cdot, s) + M \leq C W.
\end{equation}

    \item The map $s\mapsto V(\cdot, s)$ is in $C^2([0,1], \mathcal{M}_W)$.
\end{enumerate}
    
\end{as}
The bottleneck is the existence of the solutions of the rescaled Gross-Pitaevskii equation. Conditional on the existence of those solutions we can treat a more general class of external potentials.

\begin{as} \label{as: PDE input}
    There exists $\varepsilon_*>0$ such that for all $0<\varepsilon\leq \varepsilon_*$ there exists $\psi^{(\varepsilon)}\in C([0,1], \mathcal{Q}_W)$ such that $s\mapsto \langle \cdot, \psi^{(\varepsilon)}(s)\rangle_2$ is in $C^1([0,1], \mathcal{Q}_W^*)$, solves the PDE
\begin{equation}\label{eq:PDE weak epsilon}
    \begin{cases}
        i\varepsilon \frac{d}{ds} \langle u, \psi^{(\varepsilon)}(s) \rangle_2 = \int_{\mathbb{R}^3} \left(\overline{\nabla u} \cdot \nabla \psi^{(\varepsilon)}(s) + V(\cdot, s) \overline{u} \psi^{(\varepsilon)}(s) + \kappa \overline{u} \vert\psi^{(\varepsilon)}(s)\vert^2 \psi^{(\varepsilon)}(s)\right), &\text{for all } u\in \mathcal{Q}, \\
        \psi^{(\varepsilon)}(0) = \omega_0,
    \end{cases}
\end{equation}
and has a Gross-Pitaevskii energy 
\begin{equation}\label{As3GP}
    E_s = \int_{\mathbb{R}^3} \left( \vert \nabla \psi_s^{(\varepsilon)} \vert^2 + V(\cdot, s) \vert \psi_s^{(\varepsilon)} \vert^2 + \frac{\kappa}{2} \vert \psi_s^{(\varepsilon)}\vert^4 \right) dx
\end{equation}
which is absolutely continuous and satisfies for almost every $s\in [0,1]$ the balance law
\begin{equation} \label{eq: balance law}
    \frac{d}{ds} E_s = \int_{\mathbb{R}^3} (\partial_s V)(x,s) \vert \psi_s^{(\varepsilon)}(x) \vert^2 dx.
\end{equation}
\end{as}

We start with a potential which is slowly growing and slowly oscillating.

\begin{theorem} \label{thm:subquadratic}
    If Assumptions \ref{as: confining potential}, \ref{as: form} hold, and $V$ is uniformly subquadratic, i.e. $V(\cdot, s)$ is smooth and satisfies
    \begin{equation}
        \sup_{s\in [0,1]} \Vert \partial_x^\alpha V(\cdot, s) \Vert_\infty <\infty \qquad \text{for all } \vert \alpha \vert \geq 2.
    \end{equation}
    Then there exists a solution $\psi_s^{(\varepsilon)}$ of the rescaled Gross-Pitaevskii equation \eqref{GP rescaled eq} in the sense of \eqref{eq:PDE weak epsilon}. Furthermore, for every $\varepsilon_0>0$ exists a constant $D>0$ such that for all $0<\varepsilon\leq \varepsilon_0$ we have
    \begin{equation}
        \sup_{s\in [0,1]} \left\Vert e^{\frac{i}{\varepsilon} \int_0^s \mu_\tau d\tau} \psi_s^{(\varepsilon)} - \omega_s \right\Vert_{\mathcal{Q}_W} \leq D \varepsilon.
    \end{equation}
\end{theorem}

If the time-dependence is modeled by a bounded function, one can go beyond subquadratic traps.

\begin{theorem} \label{thm:poly growth}
    Let $2\leq m <\infty$. Suppose that $W\in C^\infty(\mathbb{R}^3, \mathbb{R})$ is such that $W\geq 1$ and there exist constants $c,C>0$ such that for all $x\in \mathbb{R}^3$ we have
    \begin{equation}
    \begin{split}
        c \langle x \rangle^m &\leq W(x) \leq C \langle x \rangle^m \\
        \vert \partial_x^\alpha W(x)\vert &\leq C_\alpha \langle x \rangle^{m-|\alpha|} \qquad \text{for all } \alpha\in \mathbb{N}_0^3.
    \end{split}
    \end{equation}
    Here $\langle x \rangle = \sqrt{1+x^2}$. Furthermore, let $r\in C^2([0,1], W^{1,\infty}(\mathbb{R}^3, \mathbb{R}))$ and define $V(x,s)=W(x)+r_s(x)$. Then the rescaled Gross-Pitaevskii equation admits a global solution in the sense of \eqref{eq:PDE weak epsilon} and for all $\varepsilon_0>0$ exists a constant $D>0$ such that for all $0<\varepsilon\leq \varepsilon_0$ holds
    \begin{equation}
        \sup_{s\in [0,1]} \left\Vert e^{\frac{i}{\varepsilon} \int_0^s \mu_\tau d\tau} \psi_s^{(\varepsilon)} - \omega_s \right\Vert_{\mathcal{Q}_W} \leq D \varepsilon.
    \end{equation}
\end{theorem}

Conditional on the existence of the dynamics we have again a nonlinear adiabatic theorem.
\begin{theorem} \label{thm:conditional}
    If Assumptions \ref{as: confining potential}, \ref{as: form}, \ref{as: PDE input} hold true,
    then, for every $\varepsilon_0>0$ exists a constant $D>0$ such that for all $0<\varepsilon\leq \varepsilon_0$ we have
    \begin{equation}
        \sup_{s\in [0,1]} \left\Vert e^{\frac{i}{\varepsilon}\int_0^s \mu_\tau d\tau} \psi_s^{(\varepsilon)} - \omega_s \right\Vert_{\mathcal{Q}_W} \leq D \varepsilon.
    \end{equation}
\end{theorem}

\begin{rem}
    The Gross-Pitaevskii equation gives an effective description of the dynamics of Bose-Einstein condensates. Assumption \ref{as: PDE input} is ubiquitous in this context. In particular, all current derivations of the GP equation in a scaling limit work with it, even if only implicitly so. 
\end{rem}

\subsection{An abstract setting}

First we fix a complex Hilbert space, which is the replacement of the form domain in the previous setting. Let $\mathcal{Q}\subseteq H^1(\mathbb{R}^3)$ be a dense, complex subspace which is invariant under complex conjugation. It is equipped with a norm $\Vert \cdot \Vert_\mathcal{Q}$ on $\mathcal{Q}$ which induces a complex Hilbert space structure and which is invariant under complex conjugation. We shall also use the real Hilbert space of $\mathcal{Q}_\mathbb{R}=\mathcal{Q} \cap L^2(\mathbb{R}^3, \mathbb{R})$.
\begin{as} (Form domain) \label{as: Q}
There exists a constant $C_\mathcal{Q}>0$ such that
\begin{equation}\label{eq:H1Q}
\Vert u \Vert_{H^1} \leq C_\mathcal{Q} \Vert u \Vert_\mathcal{Q}
\end{equation}
for all $u\in \mathcal{Q}$. In particular $\Vert u \Vert_{2} \leq C_\mathcal{Q} \Vert u \Vert_\mathcal{Q}$.
\end{as}
The next ingredient replaces the linear part of our problem, namely the Schrödinger operator. For every $s\in [0,1]$ let $a_s: \mathcal{Q} \times \mathcal{Q} \rightarrow \mathbb{C}$ be a closed sesquilinear form which is real in the sense that 
\begin{equation}
    a_s(\overline{u}, \overline{v}) = \overline{a_s(u,v)}
\end{equation}
for all $u,v\in \mathcal{Q}$. We will write $a_s[u]=a_s(u,u)$. Note that the domain is assumed to be independent of $s$. Let $\mathcal{Q}^*$ be the space of conjugate-linear functionals on $\mathcal{Q}$, equippe with the norm $\Vert F \Vert_{\mathcal{Q}^*}=\sup_{\Vert u \Vert_\mathcal{Q}=1} \vert F(u)\vert$. By definition, the map $f \mapsto (u\mapsto \langle u,f\rangle_2)$ is a continuous embedding $j: L^2(\mathbb{R}^3)\to \mathcal{Q}^*$; it is bounded $\Vert jf\Vert_{\caQ^*}\leq C_\mathcal{Q} \Vert u \Vert_\mathcal{Q}$ by Assumption~\ref{as: Q} and injective since $\caQ$ is dense in $L^2$. Moreover, $a_s$ is naturally seen as a bounded linear map $\mathcal{Q}\to\mathcal{Q}^*$ given by $u\mapsto a_s(\cdot,u)$. 
\begin{as} (Quadratic forms) \label{as: quadratic forms}
$a\in C^1([0, 1];\mathcal{B}(\mathcal{Q}, \mathcal{Q}^*))$ and there exist constants $c_a,C_a,m>0$ such that for all $s\in [0,1]$ and all $u\in \mathcal{Q}$ we have
\begin{equation}
    c_a \Vert u \Vert_\mathcal{Q}^2 \leq a_s[u] + m \Vert u \Vert_2^2 \leq C_a \Vert u \Vert_\mathcal{Q}^2.
\end{equation}
\end{as}
\noindent The quadratic form $a_s$ is closed and bounded below, therefore it defines a self-adjoint operator, which we will denote by $h_s$. In the previous theorems we had $h_s = -\Delta + V_s$.

\begin{as} (Regularity of ground states) \label{as: ground states}
There exists $\omega\in C^2([0, 1] ,\mathcal{Q}_\mathbb{R})$, $\mu\in C^1([0,1], \mathbb{R})$ such that
\begin{equation}\label{eq:GS of a}
    \begin{cases}
        \Vert \omega_s \Vert_2 = 1, \\
        a_s(u, \omega_s) + \kappa \langle u, \omega_s^3 \rangle_2 = \mu_s \langle u, \omega_s \rangle_2, &\text{for all } u\in \mathcal{Q},
    \end{cases}
\end{equation}
for all $s\in [0,1]$.
\end{as}
\noindent Let $S$ be the Sobolev constant $\Vert u\Vert_6\leq S\Vert\nabla u\Vert_2$ on $H^1(\R^3)$.
By Assumption~\ref{as: Q}, $\Vert u\Vert_6\leq SC_Q\Vert u\Vert_\caQ$ and $\Vert u\Vert_2\leq C_\caQ\Vert u\Vert_\caQ$, and interpolation gives
\begin{equation}\label{eq:interpolation}
\Vert u\Vert_p\le C_p\Vert u\Vert_\caQ,\qquad C_p:=C_QS^{1-\theta},\qquad u\in Q,\ p\in[2,6].
\end{equation}
where $\frac1p=\frac\theta2+\frac{1-\theta}6$. In particular $\omega_s\in L^4$ for all $s\in[0,1]$ if Assumptions~\ref{as: Q} and~\ref{as: ground states} hold. With this, the abstract GP functional is defined on $\mathcal{Q}$ by
\begin{equation}\label{eq:Nonlinear Energy}
    \caE_s(u) = a_s[u] + \frac{\kappa}{2} \Vert u\Vert_4^4
\end{equation}
and it is such that
\begin{equation}\label{eq:Abstract GS energy}
   \caE_s(\omega_s) + \frac{\kappa}{2} \Vert \omega_s\Vert_4^4 = \mu_s. 
\end{equation}

In the previous theorems, sufficient regularity of the trapping potential of the concrete Schrödinger operator implied continuity of the spectral gap. In order to state the abstract version of this, let $P_s$ be the orthogonal projection onto the span of $\omega_s$ and let $Q_s = 1-P_s$. Let moreover
\begin{equation}\label{eq:b}
b_s(u,v)= a_s(u,v) + \kappa \langle u, \omega_s^2 v \rangle_2 - \mu_s \langle u,v \rangle_2.
\end{equation}
\begin{as} (Linear spectral gap) \label{as: spectral gap}
There exists $g>0$ such that for all $u\in \mathcal{Q}$ and all $s\in [0,1]$ we have
    \begin{equation}\label{eq:b gap}
        b_s[u] \geq g \Vert Q_s u \Vert_2^2.
    \end{equation}
\end{as}
\noindent Just as $h_s$ is the operator associated with the form $a_s$, we denote by $k_s$ the linear operator associated with
\begin{equation}\label{eq:linearized operator}
    k_s = h_s + \kappa \omega_s^2 - \mu_s
\end{equation}

As discussed before, the crucial assumptions which are in general difficult to verify are the existence and regularity of the solution of the rescaled dynamics.
\begin{as} (Existence of dynamics) \label{as: existence dynamics}
There exists $\varepsilon_*>0$ such that for all $0<\varepsilon\leq \varepsilon_*$ there exists $\psi^{(\varepsilon)}\in C([0,1], \mathcal{Q})$ such that $j(\psi^{(\varepsilon)})\in C^1([0,1], \mathcal{Q}^*)$ and
\begin{equation}\label{eq:Form epsilon GP}
    \begin{cases}
        i\varepsilon \frac{d}{ds} \langle u, \psi^{(\varepsilon)}(s) \rangle_2 = a_s(u,\psi^{(\varepsilon)}(s)) + \kappa \langle \vert \psi^{(\varepsilon)}(s)\vert^2 u, \psi^{(\varepsilon)}(s)\rangle_2, &\text{for all } u\in \mathcal{Q}, \\
        \psi^{(\varepsilon)}(0) = \omega_0.
    \end{cases}
\end{equation}
\end{as}

\begin{as} (Regularity of the solution) \label{as: energy}
Denote by $a'$ the derivative of $s\mapsto a_s$. Assume that $s\mapsto \caE_s(\psi^{(\varepsilon)}(s))$ is absolutely continuous and satisfies the balance law
\begin{equation} \label{as: balance law}
    \frac{d}{ds} \caE_s (\psi^{(\varepsilon)}(s)) = a_s'[\psi^{(\varepsilon)}(s)]
\end{equation}
almost everywhere on $[0,1]$.
\end{as}

\begin{theorem} \label{thm:abstract}
    If Assumptions \ref{as: Q}--\ref{as: energy} hold, then there exist constants $0<\varepsilon_0\leq \varepsilon_*$ and $D>0$ such that

    \begin{equation}\label{eq:Main}
        \sup_{s\in [0,1]} \left\Vert e^{\frac{i}{\varepsilon}\int_0^s \mu_\tau d\tau} \psi^{(\varepsilon)}(s) - \omega_s \right\Vert_{\mathcal{Q}} \leq D \varepsilon.
    \end{equation}
for all $0<\varepsilon\leq \varepsilon_0$.
\end{theorem}
\noindent Note that the bound also holds in the more natural $H^1$ and $L^2$ norms since they are dominated by the $\mathcal{Q}$-norm, see~\eqref{eq:H1Q}. 

\begin{rem}
    While we are concentrating here on the cubic nonlinearity, our methods may also apply for different nonlinearities. In particular, a similar adiabatic theorem should hold for the Hartree equation with suitable interaction potentials. In that case however, the existence and uniqueness of the minimizer is not known in general, see \cite{frohlich2004mean}, which introduces an additional difficulty.
\end{rem}

\subsection{Outline of the paper}

We start in Section \ref{sect:abstract} with the proof of the abstract version of the nonlinear adiabatic theorem, which we have just stated. All of the results of this paper will then be proved by verifying the conditions of the abtract theorem in the various cases of interest. 

\section{Proof of Theorem \ref{thm:abstract}} \label{sect:abstract}

The proof of Theorem~\ref{thm:abstract} is fairly technical. Thus, in the first section we will give an outline of the proof, before we dive into the technical details.

\subsection{Outline of the proof}

The proof of Theorem~\ref{thm:abstract} has two steps. First, a nonlinear energy estimate
controls the component orthogonal to the ground state. Together with mass conservation, this gives closeness up to an additional phase, which the energy cannot determine. The second step controls that phase by exploiting a cancellation in the linearized equation. To lighten the notation we will suppress the superscript $\varepsilon$ on the solution and its components. All estimates are uniform for $s\in[0,1]$.

\paragraph{A time-independent reference state.}
To start with, we remove the dynamical phase and consider $\phi(s)=\exp\left(\frac{i}{\varepsilon}\int_0^s\mu_\tau\,d\tau\right)\psi(s)$, which is the vector to be compared with the instantaneous ground state $\omega_s$. Let $P_s=|\omega_s\rangle\langle\omega_s|$ and let $W(s)$ be Kato's parallel transport which is a unitary propagator in $L^2$ such that $W(s)\omega_0=\omega_s$. We define
\begin{equation*}
\zeta(s)=W(s)^*\phi(s),\qquad \zeta(0) =\omega_0.
\end{equation*}
We shall refer to this as the \emph{moving frame}: the coordinates follow $\omega_s$, so the reference state in these coordinates is the fixed vector $\omega_0 = W(s)^*\omega_s$ and
\begin{equation*}
\zeta(s)-\omega_0 = W(s)^*(\phi(s)-\omega_s)
\end{equation*}
Since $W(s)$ and its inverse are uniformly bounded on $\mathcal Q$, it suffices to estimate $\zeta-\omega_0$ in $\mathcal Q$. Unitarity on $L^2$ and mass conservation of the GP flow give $\|\zeta(s)\|_2=1$.

\paragraph{Step 1: The relative energy and the corrected Hamiltonian.}
We renormalize the GP energy functional to $\mathcal E_s(v) - \mu_s\Vert v\Vert_2^2$ and define the energy of a vector $v$ relative to that of the ground state:
\begin{equation*}
\mathcal F_s(v)
 =\mathcal E_s(W(s)v)-\mathcal E_s(W(s)\omega_0)
             -\mu_s\bigl(\|v\|_2^2-1\bigr).
\end{equation*}
In particular, $\mathcal F_s(\omega_0)=0$ for every~$s$. On normalized states it is the difference of the two energies; the
mass term makes its derivative with respect to $v$ vanish at $\omega_0$. Elementary algebra yields the equivalent expression $\mathcal F_s(v)
 =b_s[W(s)v]+\frac{\kappa}{2} \Vert |W(s)v|^2-|W(s)\omega_0|^2\Vert_2^2$ and the gap assumption~\ref{as: spectral gap} implies
\begin{equation*}
 \mathcal F_s(v)
 \geq g\|Q_0v\|_{2}^2
\end{equation*}
and a similar bound in $\mathcal Q$-norm. In other words, the functional $\mathcal F_s$ controls the norm of the leakage outside of the ground state space. In turn, the leakage bounds the variation of both terms of $\mathcal F$ and
\begin{equation*}
    \vert (\partial_s\caF_s)(v)\vert \leq C \caF_s(v).
\end{equation*}

Now $\mathcal F_s$ is not the Hamiltonian generating the evolution in the moving frame as it does not take into account the movement of the frame itself: differentiating the
changing coordinates adds a term. The correct functional is given by
\begin{equation*}
 \mathcal H_s^{(\varepsilon)}(v)
   =\mathcal F_s(v)+\varepsilon\langle v,C(s)v\rangle_2,
 \quad\text{where}\quad C(s)=-iW(s)^*W'(s),
\end{equation*}
and it is indeed such that
\begin{equation*}
    i\varepsilon\zeta'(s)=\nabla\mathcal H_s^{(\varepsilon)}(\zeta(s))
\end{equation*}
in $\mathcal Q^*$. Importantly, $\mathcal H_s^{(\varepsilon)}(\omega_0) = 0$ for all $s$ because $C$ is off-diagonal with respect to $P_0\oplus(1-P_0)$. 

The Hamiltonian evaluated along its own
flow changes only through its explicit time dependence. At the regularity
of the theorem, this is justified using the original balance law in Assumption~\ref{as: energy}:
\begin{equation*}
 \frac{d}{ds}\mathcal H_s^{(\varepsilon)}(\zeta(s))
 = (\partial_s\mathcal F_s)(\zeta(s))
       +\varepsilon\langle\zeta(s),C'(s)\zeta(s)\rangle_2.
\end{equation*}
Here, not only is the second term of order $\varepsilon$, both are actually bounded above by the relative energy $\mathcal F_s$ whenever $\|v\|_2=1$: We have just argued so for the first one, while
\begin{equation*}
    \vert \langle v,C(s)v\rangle_2\vert+|\langle v,C'(s)v\rangle_2| \leq c\mathcal F_s(v)^{1/2}
\end{equation*}
follow from the fact that $C$ is off-diagonal with respect to the time-independent splitting $P_0\oplus(1-P_0)$. This observation and the addition of a term of order $\varepsilon^2$ to $\mathcal H_s^{(\varepsilon)}(\zeta(s))$ allows for a closed differential inequality which can be integrated by Gr\"onwall to derive
\begin{equation*}
 \sup_s\mathcal F_s(\zeta(s))\le C\varepsilon^2,
\end{equation*}
since $\mathcal F_s(\zeta(0)) = \mathcal F_s(\omega_0) = 0$. In turn, this controls the leakage, namely 
\begin{equation*}
    \sup_s\|Q_0(\zeta(s)-\omega_0)\|_{\mathcal Q} =\sup_s\|Q_0\zeta(s)\|_{\mathcal Q}\le C\varepsilon.
\end{equation*}

Since $\zeta$ is normalized, smallness of $Q_0\zeta$ implies that the modulus of $\langle\omega_0,\zeta(s)\rangle_2$ is large (in fact, bounded below by $1-C\varepsilon^2$), but it fails to give any information about the phase $\exp(i \theta(s))$ of that inner product, and it is therefore insufficient to conclude anything about $\zeta - \omega_0$. A key insight to make progress is the simple observation that since $\omega_0$ is real-valued and $\theta(s)$ is expected to remain small as $\theta(0)= 0$, it suffices to control the imaginary part of $\zeta$ to control the phase. In preparation of this, we define
\begin{equation*}
    \chi(s)=e^{-i\theta(s)}\zeta(s),
\end{equation*}
for which the above gives
\begin{equation*}
    \Vert \chi(s) -\omega_0\Vert_{\mathcal{Q}}\leq C\varepsilon.
\end{equation*}

\paragraph{Step 2: Controlling the phase.}

The splitting of the problem into its real and imaginary parts is also natural from the point of view of linearization. The functional $\mathcal G_s(u)=\mathcal E_s(u)-\mu_s\|u\|_2^2$, has $\omega_s$ as its ground state and so $\nabla\mathcal G_s(\omega_s)=0$. Its linearization around $\omega_s$ gives
\begin{equation*}
\begin{split}
 \nabla\mathcal G_s(\omega_s+\delta)
   &=\mathcal L_s\delta+\mathcal R_s(\delta),\\
 \mathcal L_s(u+iv)&=(k_s+G_s)u+ik_sv,
 \qquad a,b\in\mathcal Q_{\mathbb R},
\end{split}
\end{equation*}
where $k_s=h_s+\kappa\omega_s^2-\mu_s$ and $G_s$ is multiplication by
$2\kappa\omega_s^2$. Thus, the linear part respects the real/imaginary splitting, the two sectors being connected only through the non-linearity $\caR_s$.

We apply this expansion to
$\delta=W(s)\chi(s)-\omega_s$, which is $O(\varepsilon)$ in
$\mathcal Q$ by Step~1. A simple H\"older estimate yields $\|\mathcal R_s(\delta)\|_2=O(\varepsilon^2)$. Importantly, this
uses the phase-adjusted state $\chi$, so no smallness of
$\theta$ has been assumed.

In the real space $Y=\{v\in\mathcal Q_{\mathbb R}:\langle\omega_0,v\rangle_2=0\}$, we write 
\begin{equation*}
 \chi(s) = r(s)\omega_0+\xi(s)+i\eta(s),
 \qquad \xi(s),\eta(s)\in Y.
\end{equation*}
Let $L_{\perp,s}:Y\to Y^*$ be the
real block $k_s+G_s$, transported by $W(s)^*$ and projected onto $Y$ and let similarly $f(s)=Q_0W(s)^*G_s\omega_s$. Taking the real part
of the equation in the moving frame and projecting along $\omega_0$ and onto its
orthogonal complement yields
\begin{equation*}
 -\varepsilon r(s)\theta'(s)
   =\langle f(s),\xi(s)\rangle_2+O(\varepsilon^2),
 \qquad
 -\varepsilon\eta'(s)
   =L_{\perp,s}\xi(s)+O(\varepsilon^2).
\end{equation*}
Their bounds, which we don't display here, follow from Step~1. Since $\xi=O(\varepsilon)$ and $r(s)$ is bounded away from $0$, the first equation give the preliminary $\theta'=O(1)$.

The operator $L_{\perp,s}$ is symmetric, and the spectral gap together
with $G_s\ge0$ makes it uniformly coercive. The regularity of the
coefficients then gives a vector $\alpha(s)\in Y$ satisfying
\begin{equation*}
 L_{\perp,s}\alpha(s)=f(s),
 \qquad
 \|\alpha(s)\|_{\mathcal Q}+\|\alpha'(s)\|_{\mathcal Q}\le C.
\end{equation*}
With this, testing the equation for $\eta$ against $\alpha(s)$ yields
\begin{equation*}
 -\varepsilon\langle\alpha(s),\eta'(s)\rangle_{\mathcal{Q}-\mathcal{Q}^*}+O(\varepsilon^2)
 =\langle\alpha(s),L_{\perp,s}\xi(s)\rangle_{\mathcal{Q}-\mathcal{Q}^*}
 = \langle f(s),\xi(s)\rangle_2,
\end{equation*}
which can be substituted in the first equation to give
\begin{equation*}
 \theta'(s)
   =\frac{\langle\alpha(s),\eta'(s)\rangle_2}{r(s)}+O(\varepsilon),
\end{equation*}
after cancelling one order of $\varepsilon$.

The derivative $\eta'$ need not be small, but $\eta$ itself is
$O(\varepsilon)$ because it is part of the projection of $\chi$ to the space orthogonal to $\omega_0$. We pointed out earlier that modulus $r(s)$ is of order $1$ and uniformly bounded away from $0$, while $r'(s) = O(\varepsilon)$ is provided by the imaginary part of the dynamical equation. These observations and the equation above imply that
\begin{equation*}
 \frac{d}{ds}\left(
   \theta(s)-\frac{\langle\alpha(s),\eta(s)\rangle_2}{r(s)}
 \right)=O(\varepsilon).
\end{equation*}
The quantity in parentheses vanishes initially, since
$\theta(0)=0$ and $\eta(0)=0$. Its change on $[0,1]$ is
$O(\varepsilon)$ and $|\langle\alpha,\eta\rangle_2/r|\le C\|\eta\|_{\mathcal Q}
=O(\varepsilon)$. Hence, the phase is indeed small: $\theta(s)=O(\varepsilon)$, which, when combined with the bound in Step~1, concludes the proof of~\eqref{eq:Main}.

Note that this very last argument is the analogue of the usual `integration by parts' seen so often in adiabatic theory.

\paragraph{Comparison with earlier proofs.}
Gang--Grech~\cite{GangGrech} and Fermanian-Kammerer--Joye~\cite{fermanian2020nonlinear} begin their proofs by first 
linearizing about an instantaneous stationary state. Gang--Grech use dispersive estimates and obtain $O(\varepsilon)$ in $H^2(\mathbb R^3)$, up to phase \cite[Theorem~3.1]{GangGrech}. The method is adapted to their setting: The ground state is the only bound state so scattering is the fundamental mechanism. Fermanian-Kammerer--Joye use an adiabatic
approximation of the linearized evolution and Duhamel's formula, obtaining $O(\varepsilon)$ in the underlying Hilbert-space norm, hence $L^2$ in their Schr\"odinger example \cite[Theorem~1.6 and Section~4.6]{fermanian2020nonlinear}. For them it is crucial that the time-dependent Hamiltonian is given by a time-independent reference Hamiltonian plus a bounded perturbation and that the reference Hamiltonian satisfies some strong spectral assumptions. In particular, they rule out all Schrödinger operators with confining potentials in $d\geq 2$. The allowed non-linearities depend only on a fixed, finite-dimensional projection of the wavefunction.

As we explained above, we first use the nonlinear energy to control the evolved $\psi^{(\varepsilon)}(s)$ up to phase in $\mathcal Q$. Only then do we linearize and project onto the instantaneous ground state to obtain the remaining phase estimate. The proof of Theorem~\ref{thm:abstract} therefore needs neither dispersive estimates nor spectral assumptions beyond a gap, nor any additional small parameter beyond the adabatic parameter $\varepsilon$. On the other hand, it requires the absence of any spectrum below the given eigenvalue and hence is restricted to ground state energies.

\subsection{Parallel transport and a first linearization}\label{sec:Reality if GS}

First of all, the reality of the ground state and its normalization yield the usual orthogonality
\begin{equation}\label{eq:GS orthogonality}
    \langle \omega_s, \omega_s'\rangle_2 = 0.
\end{equation}
We also recall Kato's parallel transport. For all $s\in[0,1]$, let
\begin{equation}\label{eq:K}
    K_s = [P_s',P_s].
\end{equation}
Since the ground state projection has rank one, $K_s = \vert \omega_s'\rangle\langle\omega_s\vert - \vert \omega_s\rangle\langle\omega_s'\vert$  by~\eqref{eq:GS orthogonality}, where this is meant in the sense of $\caH$. With this, $\Vert \omega_s\Vert_2 = 1$ and $\Vert\omega_s'\Vert_2\leq C_\caQ \Vert\omega_s'\Vert_\caQ$ by Assumption~\ref{as: Q},
\begin{equation}\label{eq: K bounded}
    \Vert K_s\psi\Vert_\caQ\leq 2C_\caQ \Vert\omega_s'\Vert_\caQ    \Vert\omega_s\Vert_\caQ  \Vert \psi \Vert_2,
\end{equation}
and Assumption~\ref{as: ground states} implies that $K_s\in \caB(L^2,\caQ)$ is uniformly bounded for $s\in[0,1]$. Moreover, it is in fact continuously differentiable as a map in $\caB(L^2,\caQ)$. Let now $W(s)$ be the family of unitary operators defined by $W(0) = 1$ and  
\begin{equation}\label{eq:KatoProp}
    W'(s) = K_sW(s).
\end{equation}
By the remarks above, $W(s)$ is also naturally understood as a bounded operator on $\caQ$, and $W\in C^2([0,1],\caB(\caQ))$. Importantly, $W(s)$ are intertwiners of ground state projections, see \cite{kato1950adiabatic}, namely
\begin{equation}\label{eq:parallel transport}
    W(s)P_0 = P_s W(s)
\end{equation}
for all $s\in[0,1]$. A particularly useful consequence of \eqref{eq:parallel transport} is that $W(s)\omega_0=\omega_s$.

\begin{lemma}
\label{lem:identities}
For every $s\in[0,1]$ one has $b_s(\omega_s,u)=0$ for all $u\in\caQ$, and
$W(s)^*k_sW(s)\,\omega_0=0$.
\end{lemma}
 
\begin{proof}
That $b_s(u,\omega_s)=0$ for all $u\in\caQ$ is a restatement of Assumption~\ref{as: ground states}, and it is equivalent to  $k_s\omega_s=0$ in $\caQ^*$. Since $b$ is sesquilinear, $b_s(\omega_s,u)=\overline{b_s(u,\omega_s)}=0$. Finally $W(s)^*k_sW(s)\omega_0=W(s)^*k_s\omega_s=0$ by~\eqref{eq:parallel transport}.
\end{proof}

Recalling the definition~\eqref{eq:Nonlinear Energy} of the energy functional $\caE_s$, we define
\begin{equation}
    \caG_s(v) = \caE_s(v) - \mu_s\|v\|_2^2.
\end{equation}
Then
\begin{equation*}
    \nabla \caG_s(v) = \bigl(h_s + \kappa|v|^2 - \mu_s\bigr)v,
\end{equation*}
where $\nabla \caG(v)$ is defined by $\tfrac{d}{dt} \caG(v+tu)\big|_{t=0} = 2\operatorname{Re}\langle u, \nabla \caG(v)\rangle_{\mathcal{Q}-\mathcal{Q}^*}$. It is in $\mathcal Q^*$ for any $v\in\caQ$ by Assumption~\ref{as: Q}, \ref{as: quadratic forms}. If we further let
\begin{equation}\label{eq:phi}
    \phi^{(\varepsilon)}(s) = \ep{\frac{\iu}{\varepsilon}\int_0^s\mu_\tau d\tau}\psi^{(\varepsilon)}(s),
\end{equation}
then the GP equation becomes
\begin{equation}\label{eq:GCGP}
    \mathrm{i}\varepsilon \frac{d}{ds}\ph(s) = \nabla \caG_s(\ph(s)),
\end{equation}
namely, it is a Hamiltonian system whose Hamiltonian function $\caG$ is the grand-canonical energy.

The linearization of the GP equation around the instantaneous ground state $\omega_s$ will play an essential role in the following. Let $u=\omega_s+\delta$, where $\delta\in\caQ$.  Since $\omega_s$ is real, $|u|^2=\omega_s^2+2\omega_s\operatorname{Re}\delta+|\delta|^2$, and so
\begin{equation}\label{eq:linearization}
    \nabla\caG_s(u)=(h_s+\kappa|u|^2-\mu_s)u
    =k_su+\kappa\bigl(2\omega_s\operatorname{Re}\delta+|\delta|^2\bigr)u
\end{equation}
in $\caQ^*$. Now $k_su=k_s\delta$ because $k_s\omega_s=0$, and expanding
$\kappa(2\omega_s\operatorname{Re}\delta+|\delta|^2)(\omega_s+\delta)$ yields
\begin{equation}\label{eq:linearization  physical}
  \nabla\caG_s(\omega_s+\delta)=k_s\delta+G_s\operatorname{Re}\delta+\caR_s(\delta),
\end{equation}
where $G_s$ denote the operator of multiplication by $2\kappa\omega_s^2$, and the nonlinear remainder is given by
\begin{equation*}
\caR_s(\delta)=\kappa\left(\omega_s|\delta|^2+2\omega_s(\operatorname{Re}\delta)\delta+|\delta|^2\delta\right).
\end{equation*}
For later purposes, we immediately record the following a priori estimate
\begin{equation}
\label{eq:Rbound}
  \|\caR_s(\delta)\|_2\le3\kappa\|\omega_s\|_6\|\delta\|_6^2+\kappa\|\delta\|_6^3
  \le C (\|\delta\|_\caQ^2+\|\delta\|_\caQ^3).
\end{equation}
which follows by H\"older with $\frac12=\frac16+\frac16+\frac16$ and where the second inequality is by the continuous embedding $\caQ\subset L^6$.

More importantly, (\ref{eq:linearization  physical}) together with the reality condition in Assumption~\ref{as: ground states} points to an important structure of the GP equation. The linear part $\mathcal{L} = k + G$ of~\eqref{eq:linearization  physical} is $\R$-linear but not $\C$-linear. In fact, it is block diagonal, meaning that there are $\mathbb{R}$-linear operators $L_{\pm, s}$ such that
\begin{equation*}
    \mathcal{L}_s(\xi + \iu\eta) = L_{+,s}\xi + \iu L_{-,s}\eta
\end{equation*}
for all $s\in[0,1], \xi, \eta\in \mathcal{Q}_\mathbb{R}$. It follows from the reality of $\omega_s$ that 
\begin{equation*}
    L_{-,s}\omega_s = 0,
\end{equation*}
(explicitly $L_{-,s}\omega_s = k_s\omega_s = 0$, but this is not needed), whereas
\begin{equation*}
    \langle \omega_s, L_{+,s}\omega_s\rangle_{\mathcal{Q}-\mathcal{Q}^*} = \langle \omega_s, G_s\omega_s\rangle_2 = 2\kappa\Vert \omega_s\Vert_4^4 > 0.
\end{equation*}
In the real-imaginary splitting that we shall heavily rely on, there is no movement in the imaginary direction but a strictly positive one with rate of a priori order $\varepsilon^{-1}$ in the real direction, at the linear level. Accordingly, the leakage of the $\ph$ outside of the instantaneous ground state space will best be measured by recording its phase, rather than in purely spectral terms; this will be a recurring theme of the proof.

\begin{rem}
    (i) The real-imaginary splitting is of course the origin of~\eqref{eq:GS orthogonality}. This will play an important role, not least because both $K_s$ and $W(s)$ are real operators.\\
    (ii) Furthermore, \eqref{eq:GS orthogonality} shows that this system has vanishing Berry phase. In particular, the GP equation cannot exhibit any form of topological behaviour.
\end{rem}

\subsection{Dynamics in the moving frame}\label{sec:Moving Frame}

Let now 
\begin{equation}\label{eq:nu}
    \nu(s) = W(s)^*\omega_s',
\end{equation}
which is such that $\langle \nu(s),\omega_0\rangle_2 = 0$, namely $\nu(s)\in\mathrm{Ran}(Q_0)$ for all $s\in[0,1]$. Then
\begin{equation}\label{eq:def of C}
    C(s) = -\iu W(s)^* K_s W(s)
\end{equation}
is self-adjoint and $\iu C(s) = \vert \nu(s) \rangle\langle\omega_0\vert - \vert \omega_0\rangle\langle\nu(s)\vert$. It is in particular off-diagonal with respect to the ground state splitting $\caH = P_0\caH\oplus Q_0\caH$, and so $\vert \langle \phi,C(s)\phi\rangle_2\vert\leq (\Vert C(s)P_0\phi\Vert_2 + \Vert C(s)^*P_0\phi\Vert_2)\Vert Q_0\phi\Vert_2 $. The same holds for $C'(s)$ and so
\begin{equation}\label{eq:C Bdd}
    \vert \langle v,C(s)v\rangle_2\vert + \vert \langle v,C'(s)v\rangle_2\vert
    \leq C\Vert Q_0v\Vert_2.
\end{equation}
for any normalized $v\in\caH$. Rather than studying $\phi^{(\varepsilon)}(s)$ defined in~\eqref{eq:phi} directly, we shall consider its pullback
\begin{equation}\label{eq:zeta}
    \zeta^{(\varepsilon)}(s) = W(s)^*\phi^{(\varepsilon)}(s).
\end{equation}
In the `moving frame' defined by pulling everything back with $W(s)^*$, the GP equation reads
\begin{align}
    \iu\varepsilon\frac{d}{ds}\zeta^{(\varepsilon)}(s) &= \varepsilon C(s)\zeta^{(\varepsilon)}(s) - \mu_s\zeta^{(\varepsilon)}(s) +  W(s)^*\ep{\frac{\iu}{\varepsilon}\int_0^s\mu_\tau d\tau}\iu\varepsilon\frac{d}{ds}\psi^{(\varepsilon)}(s) \nonumber\\
    &=\varepsilon C(s)\zeta^{(\varepsilon)}(s) + W(s)^*k_sW(s)\zeta^{(\varepsilon)}(s) + \caN_s(\zeta^{(\varepsilon)}(s))\label{eq: moving GP}
\end{align}
which is to be understood in $\caQ^*$ as in Assumption~\ref{as: existence dynamics}, where $k_s$ is the linearized dynamics~\eqref{eq:linearized operator} and the nonlinearity is given by
\begin{align}
    \caN_s(\zeta^{(\varepsilon)}(s)) &= \kappa W(s)^*\left(\vert \psi^{(\varepsilon)}(s)\vert^2-\omega_s^2\right)\phi^{(\varepsilon)}(s) \nonumber\\
    &= \kappa W(s)^*\left(\vert W(s)\zeta^{(\varepsilon)}(s)\vert^2-\omega_s^2\right)W(s)\zeta^{(\varepsilon)}(s).\label{eq:N}
\end{align}
Recalling the nonlinear energy functional~\eqref{eq:Nonlinear Energy}, we now define a relative energy functional on $\caQ$ by
\begin{align}
    \caF_s(v) &= \caE_s(W(s) v)-\caE_s(\omega_s) -\mu_s(\Vert v\Vert_2^2-1) \label{eq:F 1} \\
    &= \left(\caE_s(W(s) v) - \mu_s\Vert v\Vert_2^2\right) +\frac{\kappa}{2}\Vert \omega_s\Vert_4^4 \nonumber
\end{align}
where the equality is by~\eqref{eq:Abstract GS energy}. Now with $b$ as in~\eqref{eq:b}, we have that
\begin{equation} \label{eq:decomp Fs}
    \begin{split}
        \caF_s(v)  &= a[W(s) v] + \frac{\kappa}{2}\Vert W(s)v\Vert_4^4 - \mu_s\Vert W(s)v\Vert_2^2 + \frac{\kappa}{2}\Vert \omega_s\Vert_4^4 \\
    &= b_s[W_s v] - \kappa \Vert \omega_s W(s) v\Vert_2^2 + \frac{\kappa}{2}\left(\Vert W(s)v\Vert_4^4 + \Vert \omega_s\Vert_4^4\right)\\
    &=b_s[W_s v] + \frac{\kappa}{2} \left\Vert \vert W(s) v\vert^2 - \omega_s^2\right\Vert_2^2.
    \end{split}
\end{equation}

We start with some a priori bounds on the relative energy.
\begin{lemma}\label{lem:F}
    Let $g$ be as in Assumption~\ref{as: spectral gap}. There is $c>0$ such that
    \begin{equation}\label{eq:first bounds F}
        \caF_s(v)\geq c\Vert Q_0v\Vert_\caQ^2,\qquad \caF_s(v)\geq g\Vert Q_0v\Vert_2^2.
    \end{equation}
    Moreover, there is $C>0$ such that for $v\in\caQ$ with $\Vert v\Vert_2 = 1$, 
    \begin{equation}\label{eq: bound on dF}
        \vert \caF'_s(v)\vert \leq C \caF_s(v).
    \end{equation}
\end{lemma}
\begin{proof}
    By Assumption~\ref{as: spectral gap} and \eqref{eq:decomp Fs},
    \begin{equation*}
        \caF_s(v)\geq b_s[W_s v]\geq g\Vert Q_s W_s v\Vert_2^2 = g\Vert Q_0 v\Vert_2^2.
    \end{equation*}
    For the other lower bound, we first write
    \begin{equation*}
        W(s)v = \langle\omega_0,v\rangle_2 \omega_s + W(s)Q_0v,
    \end{equation*}
    and conclude that $b_s[W(s)v] = b_s[W(s)Q_0v] = b_s[Q_s W(s) v]$ by Lemma~\ref{lem:identities}. Let now $M = \sup\{\mu_s:s\in[0,1]\}$. Assumption~\ref{as: quadratic forms} yields
    \begin{align*}
        b_s[v]+ (m + M)\Vert v\Vert_2^2
        &\geq a_s[v] + \kappa\Vert \omega_s v\Vert_2^2+m\Vert u \Vert_2^2\geq c_a\Vert v\Vert_\caQ^2.
    \end{align*}
    When applied to a vector in the range of $Q_s$, Assumption~\ref{as: spectral gap} can be used again to get
    \begin{equation*}
        b_s[Q_s u]\geq \left(1+\frac{m+M}{g}\right)^{-1}c_a \Vert Q_s u\Vert_\caQ^2.
    \end{equation*}
    Summarizing, we obtain
    \begin{equation*}
        \caF_s(v)\geq b_s[W(s) v] = b_s[Q_s W(s)v] \geq \frac{c_ag}{g+(m+M)}\Vert Q_s W(s) u\Vert_\caQ^2.
    \end{equation*}
    Finally, $\Vert Q_s W(s) u\Vert _\caQ = \Vert  W(s)Q_0 u\Vert_\caQ\geq C_{W^{-1}}\Vert  Q_0 u\Vert_\caQ $ since $W(s)^{-1}$ is bounded on $\caQ$. This is the first lower bound with
    \begin{equation*}
        c=\frac{c_ag}{C_{W^{-1}}(g+(m+M))}.
    \end{equation*}
    
    It remains to prove the bound on the derivative. We will bound the derivatives of the two terms in \eqref{eq:decomp Fs} separately. Since $\omega_s$ is a ground state, $b_s(u,\omega_s) = 0$ for $s\in[0,1]$ and all $u\in\caQ$ and so $\frac{d}{ds} b_s[W_s v]= \frac{d}{ds} b_s[Q_sv]$. Since $\partial_s b_s(u,v) = a'_s(u,v) + 2\kappa\langle u,\omega_s\omega_s'v\rangle_2 - \mu'_s\langle u,v\rangle_2$, Assumptions~\ref{as: Q}-\ref{as: ground states} give that $b\in C^1([0,1];\caB(\caQ,\caQ^*))$ and we have that
    \begin{equation}\label{first bound}
        \vert \frac{d}{ds} b_s[W_s v]\vert\leq C\Vert Q_0 v\Vert_\caQ^2\leq C\caF_s(v)
    \end{equation}
    by the bound just proven.
    
    Next we estimate the derivative of the second term in \eqref{eq:decomp Fs}. We denote $\rho(s) = \vert W(s) v\vert^2 - \omega_s^2$ and let $\theta$ be the argument of $\langle \omega_0,v\rangle_2$ (we pick an arbitrary one if the scalar product vanishes). Then $\delta(s) = W(s) v - \ep{\iu\theta}\omega_s$ is such that
    \begin{equation*}
        P_0W(s)^*\delta(s) = P_0 v-\ep{\iu\theta}\omega_0 = 0,\qquad Q_0W(s)^*\delta(s) = Q_0v.
    \end{equation*}
    Hence by unitary invariance of the $L^2$ norm,
    \begin{equation*}
        \Vert \delta(s)\Vert_2 = \Vert W(s)^*\delta(s)\Vert_2
        =\Vert Q_0 v\Vert_2.
    \end{equation*}
    In $\caQ$, the $L^2$ adjoint $W(s)^*$ in not unitary but it is bounded with bounded inverse and so
    \begin{equation*}
        \Vert \delta(s)\Vert_\caQ \leq C_{(W^*)^{-1}}\Vert W(s)^* \delta(s)\Vert_\caQ = C_{(W^*)^{-1}}\Vert Q_0 v\Vert_\caQ.
    \end{equation*}
    Now,
    \begin{equation*}
        \partial_s\rho(s) = 2\mathrm{Re}(\overline{W(s)v}K_sW(s)v) - 2\omega_s K_s\omega_s
        =2\mathrm{Re}\left(\overline{\delta(s)}K_sW(s)v + \ep{-\iu\theta}\omega_s K_s\delta(s)\right).
    \end{equation*}
    Now, $K$ is uniformly bounded from $L^2$ to $\caQ$, see~\eqref{eq: K bounded}, and since $\caQ$ is continuously embedded in $L^4$, see~\eqref{eq:interpolation} we conclude that $\Vert K_s v\Vert_4\leq C \Vert v\Vert_\caQ$ and $\Vert K_s v\Vert_4\leq C\Vert v\Vert_2$. Altogether,
    \begin{align*}
        \Vert \partial_s\rho(s)\Vert_2
        &\leq 2\Vert \delta(s)\Vert_4\Vert K_sW(s)v\Vert_4 + 2\Vert \omega_s\Vert_4\Vert K_s\delta(s)\Vert_4 \\
        &\leq C\left(\Vert \delta(s)\Vert_\caQ + \Vert \omega_s\Vert_4 \Vert \delta(s)\Vert_\caQ \right)\leq C\Vert Q_0 v\Vert_\caQ
    \end{align*}
    uniformly in $s$, where we also used that $\Vert W(s)v\Vert_2 = \Vert v\Vert_2 = 1$. Hence,
    \begin{align*}
        \left\vert \partial\Vert\rho(s)\Vert_2^2\right\vert
        \leq 2\Vert \rho\Vert_2\Vert \partial_s\rho\Vert_2
        \leq C \caF_s(v)^{1/2}\Vert Q_0 v\Vert_\caQ\leq C\caF_s(v)
    \end{align*}
    by~\eqref{eq:first bounds F}, and we used $\Vert \rho\Vert_2^2\leq (1+C)\caF_s(v)$ by~\eqref{eq:decomp Fs}. This and~\eqref{first bound} conclude the proof of~\eqref{eq: bound on dF}. 
\end{proof}

We now express the GP equation~\eqref{eq:GCGP} in the moving frame. Recalling that $\zeta^{(\varepsilon)} = W^*\phi^{(\varepsilon)}$, we have that $\mathrm{i}\varepsilon(\ph)' = \mathrm{i}\varepsilon K W\ze + W\mathrm{i}\varepsilon(\ze)'$, so
\begin{align*}
    \mathrm{i}\varepsilon \frac{d}{ds}\zeta^{(\varepsilon)}(s) 
    &= W(s)^*(\nabla \caG_s)(W(s)\zeta^{(\varepsilon)}(s))-\mathrm{i}\varepsilon\, W(s)^*K_sW(s)\zeta^{(\varepsilon)}(s) \\
    &= \nabla\big(\caG_s(W_s\zeta^{(\varepsilon)}(s))\big) + \varepsilon C(s)\zeta^{(\varepsilon)}(s).
\end{align*}
The second term is the `Coriolis-type' term generated by the frame's own motion. And since $C(s)$ is self-adjoint, $\varepsilon C\zeta = \varepsilon\nabla \langle\zeta, C\zeta\rangle_2$. We conclude that, as an element of $\caQ^*$,
\begin{equation*}
    \mathrm{i}\varepsilon\frac{d}{ds}\zeta^{(\varepsilon)}(s)  = \nabla\caH^{(\varepsilon)}_{s}(\zeta^{(\varepsilon)}(s)),
\end{equation*}
where
\begin{equation}\label{eq: Moving H}
    \caH^{(\varepsilon)}_{s}(v) = \caF_s(v) + \varepsilon\langle v, C(s)v\rangle_2.
\end{equation}
In other words, $\caH^{(\varepsilon)}_{s}$ is the Hamiltonian in the moving frame.
 
Let $G_s$ denote the operator of multiplication by $2\kappa\omega_s^2$ and put
\begin{equation}
\label{eq:Gtilde}
  \Gt_s:=W(s)^*G_sW(s).
\end{equation}
Both are nonnegative and symmetric, and by~\eqref{eq:interpolation} with $p=6$,
\begin{equation}
\label{eq:Gbound}
  \|G_su\|_2\le2\kappa\|\omega_s\|_6^2\|u\|_6\le C\|u\|_\caQ,
  \qquad
  \|\Gt_s\omega_0\|_2=2\kappa\|\omega_s\|_6^3\le C,
\end{equation}
uniformly in~$s$.
 
\begin{lemma}
\label{lem:mass}
Under Assumption~\ref{as: existence dynamics}, $\|\psi^{(\varepsilon)}(s)\|_2=1$ for all $s\in[0,1]$, and consequently
$\|\phi^{(\varepsilon)}(s)\|_2=\|\zeta^{(\varepsilon)}(s)\|_2=1$.
\end{lemma}
 
\begin{proof}
Since $\ps\in C([0,1];\caQ)$ and $j(\ps)\in C^1([0,1];\caQ^*)$, the Lions-Magenes lemma in
the triple $\caQ\subset L^2\subset\caQ^*$ (see \cite[Chapter III, §1, Lemma 1.2]{Temam1979}) shows that $s\mapsto\|\ps(s)\|_2^2$ is
absolutely continuous with $\frac{d}{ds}\|\ps\|_2^2=2\operatorname{Re}\langle\ps,(\ps)'\rangle_{\mathcal{Q}-\mathcal{Q}^*}$.
Taking $u=\ps_s$ in~\eqref{eq:Form epsilon GP} at fixed $s$,
\[
  \mathrm{i}\varepsilon\,\langle\ps_s, (\ps_s)'\rangle_{\mathcal{Q}-\mathcal{Q}^*}=a_s[\ps_s]+\kappa\|\ps_s\|_4^4\in\mathbb{R}.
\]
Hence $\langle\ps,(\ps)'\rangle_{\mathcal{Q}-\mathcal{Q}^*}$ is purely imaginary  and so $\frac{d}{ds}\|\ps\|_2^2=0$, which concludes the proof since $\ps(0) = \omega_0$ is normalized.
\end{proof}
 
\begin{rem}
\label{rem:F-along}
By Lemma \ref{lem:mass} the constraint term in~\eqref{eq:F 1} vanishes identically along
the solution, so
\begin{equation}
\label{eq:F-along}
  \caF_s(\ze(s))=\caE_s(\ph(s))-\caE_s(\omega_s),
\end{equation}
for all $s\in[0,1]$ and in particular $\caF_0(\ze(0))=\caF_0(\omega_0)=0$.  Moreover, \eqref{eq:C Bdd} and \eqref{eq: bound on dF} are applicable at $v=\ze(s)$.
\end{rem}
 
\subsection{The corrected balance law} \label{sec:corrected-balance-law}
 
The functional $\Ham_s$~\eqref{eq: Moving H} was identified as the Hamiltonian of the flow in the moving frame. We now turn to its evolution along the flow. Note that it cannot be obtained by a formal chain rule: $\ze$ is differentiable only in $\caQ^*$, and we do not assert that the cubic term preserves any operator domain. However, we can rely on the balance law~\eqref{eq: balance law}. 
 
\begin{lemma}
\label{lem:balance}
Let $\ze$ be as in~\eqref{eq:zeta}. Then $s\mapsto\Ham_s(\ze(s))$ is absolutely
continuous on $[0,1]$ and, almost everywhere,
\begin{equation}
\label{eq:balance}
  \frac{d}{ds}\,\Ham_s\bigl(\ze(s)\bigr)
  =\caF_s'\bigl(\ze(s)\bigr)+\varepsilon\,\langle\ze(s),C'(s)\ze(s)\rangle_2.
\end{equation}
\end{lemma}

\begin{proof}
We first consider the derivative of the relative energy. By \eqref{eq:F-along} it suffices to
differentiate $\caE_s(\ph(s))-\caE_s(\omega_s)$.  The GP energy is gauge
invariant, $\caE_s(\ph(s))=\caE_s(\ps(s))$, so Assumption~\ref{as: energy} gives
\begin{equation*}
    \frac{d}{ds}\caE_s(\ph(s))=a_s'[\ph(s)]
\end{equation*}
almost everywhere.  For the stationary
term, $\omega\in C^2([0,1];\caQ_\R)$ by Assumption~\eqref{as: ground states} and hence
\begin{equation*}
      \frac{d}{ds}\caE_s(\omega_s)
  =a_s'[\omega_s]+2\operatorname{Re}\left(a_s(\omega_s',\omega_s)+\kappa\langle\omega_s',\omega_s^3\rangle_2\right)
  =a_s'[\omega_s]+2\operatorname{Re}\left(\mu_s\langle\omega_s',\omega_s\rangle_2\right)
  =a_s'[\omega_s],
\end{equation*}
by~\eqref{eq:GS of a} with $u=\omega_s'$ and the orthogonality~\eqref{eq:GS orthogonality}.  Hence
\begin{equation}
\label{eq:step1}
  \frac{d}{ds}\caF_s(\ze(s))=a_s'[\ph(s)]-a_s'[\omega_s].
\end{equation}
 
\noindent We turn to the derivative of the $C(s)$ correction. Recalling~\eqref{eq:def of C},
\begin{align*}
  \langle\ze(s), C(s)\ze(s)\rangle_2 &=-\iu\langle\ph(s) K_s\ph(s)\rangle_2\\
  &=-\iu\langle \ph(s),\omega_s'\rangle_2\langle \omega_s,\ph(s)\rangle_2
  +\iu \langle\ph(s),\omega_s\rangle_2\langle\omega_s',\ph(s)\rangle_2,
\end{align*}
which is continuous by Assumption~\ref{as: ground states}. Moreover,
$j(\ph)\in C^1([0,1];\caQ^*)$ by Assumption~\ref{as: existence dynamics}, so the $\caQ$--$\caQ^*$ product rule applies.  Since
$-\iu K_s$ is self-adjoint,
\begin{equation*}
     -\iu\frac{d}{ds}\langle\ph(s),K_s\ph(s)\rangle_2
     =-\iu\langle\ph(s),K_s'\ph(s)\rangle_2+2\operatorname{Re}\langle\ph(s),-\iu K_s(\ph)'(s)\rangle_2, 
\end{equation*}
and since $(\ph)'=\frac{1}{\iu\varepsilon}\nabla\caG(\ph)$,
\begin{equation*}
    2\operatorname{Re}\langle\ph(s),-\iu K_s(\ph(s))'\rangle_2
    =\frac{2}{\varepsilon}\operatorname{Re}\langle\ph(s),-K_s\nabla\caG_s(\ph(s))\rangle_{\mathcal{Q}-\mathcal{Q}^*}
    =\tfrac{2}{\varepsilon}\operatorname{Re}\langle K_s\ph(s),\nabla\caG_s(\ph(s))\rangle_{\mathcal{Q}-\mathcal{Q}^*}.
\end{equation*}
Note that the last pairing between $\caQ$ and $\caQ^*$ is legitimate because
$K_s\in\mathcal B(\caH,\caQ)$ as established at the very beginning of Section~\ref{sec:Moving Frame}. Hence,
\begin{equation}
\label{eq:step2}
  \frac{d}{ds}\Big(\varepsilon\langle\ze(s),C(s)\ze(s)\rangle_2\Big)
  =-\mathrm{i}\varepsilon\langle\ph(s),K_s'\ph(s)\rangle_2+2\operatorname{Re}\langle K_s\ph(s),\nabla\caG_s(\ph(s))\rangle_{\mathcal{Q}-\mathcal{Q}^*}.
\end{equation}

It remains to compute the right hand side of~\eqref{eq:balance}. Differentiating~\eqref{eq:F 1} in $s$ at fixed $v$ and using $\partial_s(W v)=KWv$ yields
\begin{equation}
\label{eq:step3}
  \caF_s'(v)=a_s'[W(s)v]-a_s'[\omega_s]+2\operatorname{Re}\bigl\langle K_sW(s)v,\nabla\caG_s(W(s)v)\bigr\rangle_{\mathcal{Q}-\mathcal{Q}^*} ,
\end{equation}
since the term $-\mu_s'(\|v\|_2^2-1)$ vanishing on the mass shell $\|v\|_2=1$. Differentiating now~\eqref{eq:def of C}
\begin{equation*}
    C'(s)=-\mathrm{i}W(s)^*\bigl(-K_sK_s+K_s'+K_sK_s\bigr)W(s)=-\mathrm{i}W(s)^*K_s'W(s),
\end{equation*}
so that $\langle\ze,C'\ze\rangle_2=-\iu\langle\ph,K'\ph\rangle_2$. Adding~\eqref{eq:step1} and~\eqref{eq:step2} and comparing with~\eqref{eq:step3} at~$v=\ze$, where $W\ze=\ph$, the two occurrences of $2\operatorname{Re}\langle K\ph,\nabla\caG(\ph)\rangle_{\mathcal{Q}-\mathcal{Q}^*}$ cancel and \eqref{eq:balance} follows.
\end{proof}
 
\begin{rem}
The content of \eqref{eq:balance} is the usual fact that a Hamiltonian evaluated along its own flow changes only through its explicit time dependence; the proof above is just the rigorous version of that formal statement at the regularity of Assumptions~\ref{as: existence dynamics}, \ref{as: energy}.
\end{rem}
 
\subsection{A priori estimates}

We turn to two a priori estimates in the moving frame: one for the relative energy along the flow, and the second one estimating in norm the leakage outside of the ground state space. 
\begin{prop}
\label{prop:apriori}
There are $\varepsilon_0\in(0,\varepsilon_*]$ and $C<\infty$, independent of $\varepsilon$, such that
for all $0<\varepsilon\le\varepsilon_0$,
\begin{equation}
\label{eq:apriori}
  \sup_{s\in[0,1]}\caF_s(\ze(s))\le C\varepsilon^2,
  \qquad
  \sup_{s\in[0,1]}\|Q_0\ze(s)\|_\caQ\le C\varepsilon .
\end{equation}
\end{prop}
 
\begin{proof}
By Lemma \ref{lem:mass} we may use~\eqref{eq:C Bdd} and the second bound of Lemma~\ref{lem:F} at
$v=\ze$ to get
\begin{equation}
\label{eq:C-by-F}
  |\langle\ze(s),C(s)\ze(s)\rangle_2|+|\langle\ze(s),C'(s)\ze(s)\rangle_2|
  \leq C\|Q_0\ze(s)\|_2\leq \frac{C}{\sqrt g}\caF_s(\ze(s))^{1/2}.
\end{equation}
We let $A_0=1+C^2/(2 g)$ and define
\begin{equation*}
  \Gamma^{(\varepsilon)}(s)=\Ham_s(\ze(s))+A_0\varepsilon^2
  =\caF_s(\ze(s))+\varepsilon\langle\ze(s),C(s)\ze(s)\rangle_2+A_0\varepsilon^2 .
\end{equation*}
By~\eqref{eq:C-by-F},
\begin{align*}
    \Gamma^{(\varepsilon)}(s)
    \geq \caF_s(\ze(s)) - \frac{\varepsilon C}{\sqrt g}\caF_s(\ze(s))^{1/2} + A_0\varepsilon^2,
\end{align*}
and Young's inequality $\varepsilon C g^{-1/2}\caF^{1/2}\le\frac12\caF+C^2/(2g)\varepsilon^2$ yields
\begin{equation}
\label{eq:Gamma-lower}
  \Gamma^{(\varepsilon)}(s)\geq \frac{1}{2}\caF_s(\ze(s))+\varepsilon^2>0.
\end{equation}
By Lemma \ref{lem:balance}, \eqref{eq: bound on dF} and \eqref{eq:C-by-F},
\[
  \bigl(\Gamma^{(\varepsilon)}\bigr)'(s)=\caF_s'(\ze(s))+\varepsilon\langle\ze,C'(s)\ze\rangle_2
  \le C\caF_s(\ze(s))+\frac{\varepsilon C}{\sqrt g}\caF_s(\ze(s))^{1/2}\le C\Gamma^{(\varepsilon)}(s),
\]
where the last step used $\caF(\ze)\le2\Gamma^{(\varepsilon)}$ and Young's inequality again.
 
At $s=0$, $\ze(0)=W(0)^*\ph(0)=\omega_0$, so $\caF_0(\ze(0))=0$ by
Remark~\ref{rem:F-along}, while $\mathrm{i}C(0)\omega_0=\nu(0)$, see~\eqref{eq:nu}, and
$\langle\omega_0,\nu(0)\rangle_2=0$ reads $\langle\omega_0,C(0)\omega_0\rangle_2 =0$.  Hence
$\Gamma^{(\varepsilon)}(0)=A_0\varepsilon^2$, and Gr\"onwall's inequality yields
\begin{equation*}
    \Gamma^{(\varepsilon)}(s)\leq C\varepsilon^2
\end{equation*}
uniformly in $[0,1]$. The first bound in~\eqref{eq:apriori} then follows from~\eqref{eq:Gamma-lower}, the second from the first bound of~\eqref{eq:first bounds F}.
\end{proof}

\subsection{Polar decomposition in the moving frame}
 
By Proposition~\ref{prop:apriori} and Lemma~\ref{lem:F}, $\|Q_0\ze\|_2\le g^{-1/2}\caF(\ze)^{1/2}\le C\varepsilon$, so
\begin{equation*}
    \vert\langle\omega_0,\ze(s)\rangle_2\vert^2=1-\|Q_0\ze(s)\|_2^2\ge1-C\varepsilon^2.
\end{equation*}
By possibly shrinking $\varepsilon_0$, we may therefore assume
\begin{equation}
\label{eq:rbounds}
  \re(s)=\vert\langle \omega_0,\ze(s)\rangle_2\vert\geq \tfrac12,
  \quad\text{and}\quad
  \vert \re(s)-1\vert \leq C\varepsilon^2.
\end{equation}

With this, we turn to the argument $\thee(s)$ of $\langle\omega_0,\ze(s)\rangle_2$. Since $s\mapsto\langle\omega_0,\ze(s)\rangle_2$ is $C^1$ and nonvanishing, $s\mapsto\thee(s)$ may be chosen $C^1$, and $\thee(0)=0$ because $\ze(0)=\omega_0$. Set
\begin{equation}
    \be(s)=\varepsilon(\thee)'(s).
\end{equation}

\begin{lemma}\label{lem:small beta}
    There is $C>0$ such that 
    \begin{equation} \label{eq:small beta}
        |\be(s)|\le C\varepsilon.
\end{equation}
for all $s\in[0,1]$.
\end{lemma}

\begin{rem}
\label{rem:crude}
This estimate gives only $\theta'=\beta/\varepsilon=\caO(1)$, hence
$\theta=\caO(1)$.  This is not enough, and we provide a better one in the final step.
\end{rem}

\begin{proof}
Let
\begin{equation}
\label{eq:polar}
  \chie(s)=\mathrm{e}^{-\mathrm{i}\thee(s)}\ze(s),
\end{equation}
which is real, and decompose
\begin{equation}\label{eq:decomp of chi}
    \chie(s)= \re(s)\omega_0+\left(\xie(s)+\iu\et(s)\right),
\end{equation}
where $\xie,\et\in Y:=Q_0\caQ_\R$. By Proposition \ref{prop:apriori},
\begin{equation}
\label{eq:xieta}
  \Vert\xie(s)\Vert_\caQ+\Vert\et(s)\Vert_\caQ\leq 2\Vert Q_0 \chie(s) \Vert_\mathcal{Q}\leq  C\varepsilon .
\end{equation}
Here we have used that the norm on $\mathcal{Q}$ is invariant under complex conjugation in order to estimate the norm of the real/imaginary part by the full norm.
 
In order to estimate $\be$, we turn to the GP equation, and we aim to decompose it into a real and imaginary part. Recalling ~\eqref{eq:N}, we see that $\caN_s(\mathrm{e}^{\mathrm{i}\theta}\chi)=\mathrm{e}^{\mathrm{i}\theta}\caN_s(\chi)$, and so the moving-frame equation~\eqref{eq: moving GP} written in terms of $\chie$ reads
\begin{equation}
\label{eq:chi-eq}
  \iu\varepsilon(\chie)'(s)=W(s)^*k_sW(s)\chie(s)+\caN_s(\chie(s))+\varepsilon C(s)\chie(s)+\be(s)\chie(s).
\end{equation}
Put
\begin{equation}
\label{eq:delta}
  \de(s)=W(s)\chie(s)-\omega_s=W(s)\bigl(\chie(s)-\omega_0\bigr),
\end{equation}
which is $\mathrm{e}^{-\iu\thee}$ times the quantity denoted $\delta$ in the proof of Lemma~\ref{lem:F} at $v=\ze$, and in particular has the same norms. We claim that
\begin{equation}
\label{eq:N-expanded}
  \caN_s(\chie(s))=\Gt_s\bigl((\re(s)-1)\omega_0+\xie(s)\bigr)+\Rt(s),
  \quad\text{where}\quad \Rt(s)=W(s)^*\caR_s(\de(s)),
\end{equation}
which we will show at the end of the proof. By \eqref{eq:rbounds}, \eqref{eq:xieta} and the boundedness of $W(s)$ on $\caQ$ we
have $\|\de\|_\caQ\le C\varepsilon$, hence by \eqref{eq:Rbound}
\begin{equation}
\label{eq:Rsmall}
  \|\Rt(s)\|_2\le C\varepsilon^2 .
\end{equation}
 
The operator $\mathrm{i}C(s)=|\nu(s)\rangle\langle\omega_0|-|\omega_0\rangle\langle\nu(s)|$
is real and antisymmetric, and acts by
\begin{equation}
\label{eq:iCaction}
  \mathrm{i}C(s)\,\omega_0=\nu(s),
  \qquad
  \mathrm{i}C(s)v=-\langle\nu(s),v\rangle_2\,\omega_0\quad(v\in Y),
\end{equation}
since $\langle\nu(s),\omega_0\rangle_2=\langle\omega'_s,\omega_s\rangle_2 = 0$.  Writing
$\varepsilon C\chie=-\iu\varepsilon \iu C(\re\omega_0+\xie)+\varepsilon \iu C(s)\et$
and inserting \eqref{eq:N-expanded} into \eqref{eq:chi-eq}, the real and
imaginary parts give the system (we temporarily suppress the $s$-dependence and  $(\cdot)^{(\varepsilon)}$ for notational clarity)
\begin{align}
\label{eq:real}
  -\varepsilon\eta'&=W^*kW \xi+\Gt\bigl((r-1)\omega_0+\xi\bigr)
    +\operatorname{Re}\Rt-\varepsilon\langle\nu,\eta\rangle_2\omega_0+\beta(r\omega_0+\xi),\\
\label{eq:imag}
  \varepsilon(r'\omega_0+\xi')&=W^*kW\eta+\operatorname{Im}\Rt
    -\varepsilon(r\nu-\langle\nu,\xi\rangle_2 \omega_0)+\beta\eta,
\end{align}
since $W$ is a real operator and $W^*kW\omega_0=0$. Taking the $L^2$-inner product with $\omega_0$ and using that $\langle\omega_0,\xie\rangle_2=\langle\omega_0,\et\rangle_2=0$ for all $s\in[0,1]$ and hence also for the derivatives, as well as $\langle\omega_0,\nu\rangle_2=0$ again, we conclude that
\begin{align}
\label{eq:pair-real}
  0&=\langle\omega_0,\Gt_s\xie(s)\rangle_2+\lambda(s)+\be(s)\re(s),\\
\label{eq:pair-imag}
\varepsilon(\re)'&=\langle\omega_0,\operatorname{Im}\Rt(s)\rangle_2+\varepsilon\langle\nu(s),\xie(s)\rangle_2,
\end{align}
where $\lambda=(\re-1)\langle\omega_0,\Gt\omega_0\rangle_2+\langle\omega_0,\operatorname{Re}\Rt\rangle_2-\varepsilon\langle\nu,\et\rangle_2$. By~\eqref{eq:rbounds}, $\vert \re-1\vert=\caO(\varepsilon^2)$, by~\eqref{eq:Rsmall} $\Vert\Rt\Vert_2 = \caO(\varepsilon^2)$ and $\Vert\et\Vert_2 = \caO(\varepsilon)$ by~\eqref{eq:xieta}, so that
\begin{equation}\label{eq:lambda}
    |\lambda(s)|\le C\varepsilon^2.
\end{equation} 
Moreover, $\re\geq \frac{1}{2}$  by~\eqref{eq:pair-real} and $\vert\langle\omega_0,\Gt\xie\rangle_2\vert=\vert\langle\Gt\omega_0,\xie\rangle_2\vert\leq\|\Gt\omega_0\|_2\|\xie\|_2 = \caO(\varepsilon)$ by~\eqref{eq:xieta} imply that
\begin{equation*}
    |\beta(s)|\le C\varepsilon.
\end{equation*}
Finally, we record an inequality which will prove useful later on. Inequality \eqref{eq:Rsmall} and $\Vert\xie\Vert_2 = \caO(\varepsilon)$ by~\eqref{eq:xieta} yield
\begin{equation}\label{eq:r'}
  \vert (\re)'(s)\vert \leq C\varepsilon.
\end{equation}  

To conclude the proof of the lemma, it remains to justify~\eqref{eq:N-expanded}. Observe first that $\caN$ is exactly the nonlinear part of the
pulled-back gradient. Indeed,
\begin{equation*}
    \nabla\caG_s(u)=(h_s+\kappa|u|^2-\mu_s)u =k_su+\kappa\bigl(|u|^2-\omega_s^2\bigr)u ,
\end{equation*}
and applying $W^*$ at $u=Wv$ gives, for every $v\in\caQ$,
\begin{equation}
\label{eq:N-identity}
  W(s)^*\nabla\caG_s\bigl(W(s)v\bigr)=W(s)^*k_sW(s)\,v+\caN_s(v),
\end{equation}
by the definition~\eqref{eq:N} of $\caN_s$.  Now take $v=\chie$.  By
\eqref{eq:delta} we have $W\chie=\omega_s+\de$, so expanding the left-hand
side of \eqref{eq:N-identity} with \eqref{eq:linearization} yields
\begin{equation*}
    W(s)^*k_sW(s)\chie+\caN_s(\chie)
    =W(s)^*k_s\de+W(s)^*G_s\operatorname{Re}\de+W(s)^*\caR_s(\de).
\end{equation*}
The two $k_s$ terms cancel: by \eqref{eq:delta} and Lemma \ref{lem:identities},
\begin{equation*}
  W(s)^*k_s\de=W(s)^*k_sW(s)(\chie-\omega_0)
  =W(s)^*k_sW(s)\chie-W(s)^*k_sW(s)\omega_0,
\end{equation*}
and the second term vanishes. For the middle term, we recall that $W(s)$ is real and so
\begin{equation*}
    \operatorname{Re}\de(s)=W(s)\operatorname{Re}(\chie(s)-\omega_0) =W(s)\left((\re(s)-1)\omega_0+\xie(s)\right)
\end{equation*}
by definition, see~\eqref{eq:decomp of chi}, and because all these functions are real.  Therefore $W^*G\operatorname{Re}\de=\Gt\bigl((\re-1)\omega_0+\xie\bigr)$ by
\eqref{eq:Gtilde}, and \eqref{eq:N-expanded} follows.
\end{proof}

\subsection{An improved phase estimate}
 
On the real Hilbert space $Y=Q_0\caQ_\R$ we define the quadratic form
\begin{equation}
\label{eq:ell}
  \ell_s(v,u):=b_s\bigl(W(s)v,W(s)u\bigr)+\langle v, \Gt_s u\rangle_2.
  \qquad  .
\end{equation}
Let $L_{\perp,s}:Y\longrightarrow Y^*$ be the corresponding operator, namely
\begin{equation*}
    \langle v, L_{\perp,s}u\rangle_{\mathcal{Q}-\mathcal{Q}^*} = \ell_s(v,u)
\end{equation*}
for any $u,v \in Y$, where the left hand side is $\caQ_\R-\caQ_\R^*$ pairing. In other words, $L_{\perp,s} = j(Q_0 W(s)^*(k_s + G_s)W(s)Q_0)$ acting on the real space $\caQ_\R$.

By Assumptions~\ref{as: Q},\ref{as: quadratic forms} and the boundedness of $W(s)$ on $\caQ$, the forms $\ell_s$ are uniformly bounded. They are uniformly coercive too, as in the proof of Lemma~\ref{lem:F}. Indeed, we already showed that $b_s[W(s)v]\geq c\Vert Q_0v\Vert _\caQ^2$ for all $v\in\caQ$. If $v\in Y$, then $Q_0v=v$. Finally, we also have $\langle v,\Gt_sv\rangle_2=2\kappa\Vert \omega_s(W(s) v\Vert_2^2\ge0$, and so
$\ell_s(w,w)\ge c\|w\|_\caQ^2$ uniformly in $s$.
\begin{lemma}
\label{lem:corrector}
For all $s\in[0,1]$, the operator $L_{\perp,s}$ is an isomorphism of $Y$ onto $Y^*$ and
$\sup_{s\in[0,1]}\|L_{\perp,s}^{-1}\|_{Y^*\to Y}<\infty$. There is a unique $\alpha(s)\in Y$ with $\ell_s(v,\alpha(s))=\langle v,\Gt_s\omega_0\rangle_2$ for every $v\in Y$ and
\begin{equation}
\label{eq:alpha}
  \alpha(s)=L_{\perp,s}^{-1}Q_0\Gt_s\omega_0.
\end{equation}
Moreover, $\alpha\in C^1([0,1];Y)$, with
\begin{equation}
\label{eq:alpha-C1}
  \alpha'(s)=L_{\perp,s}^{-1}\Bigl(\bigl(Q_0\Gt_s\omega_0\bigr)'-L_{\perp,s}'\alpha(s)\Bigr).
\end{equation}
\end{lemma}
 
\begin{proof}
Since $\ell_s$ is bounded and coercive, the first part of the theorem is the Lions-Lax-Milgram theorem, and~\eqref{eq:alpha} is the definition of the weak inverse. The map $s\mapsto L_{\perp,s}\in\mathcal B(Y,Y^*)$ is $C^1$: the derivative of
$b_s(W(s)\cdot,W(s)\cdot)$ involves $a_s'$, $\omega_s'$, $\mu_s'$ and $W'=K_sW$, all of which are controlled by Assumptions~\ref{as: quadratic forms} and~\ref{as: ground states}. Recalling the definition \eqref{eq:Gtilde} of $\Gt$, we write $\langle v,\Gt u\rangle_2 = 2\kappa\int\omega_2\overline{Wv}Wu$. Since $W(s)$ is real and $\partial_s W(s) v = K_sW(s)v$,
\begin{equation*}
    \partial_s\langle v, \widetilde{G}_s u\rangle_2 = 2\kappa\int_{\mathbb{R}^3}\left(
    2\omega_s\omega_s'(W(s)v)(W(s)u) + \omega_s^2 (K_sW(s)v)(W(s)u) + \omega_s^2 (W(s)v)(K_sW(s)u)
    \right)
\end{equation*}
and $\Vert KWv\Vert_\caQ\leq C\Vert v\Vert_2\leq C\Vert v\Vert _\caQ$, we conclude that $\vert \partial_s\langle v, \widetilde{G}_s u\rangle_2 \vert\leq C\Vert v\Vert_\caQ\Vert u\Vert_\caQ$ by Assumptions~\ref{as: Q},\ref{as: ground states}. Likewise, writing $\langle v,\widetilde{G} \omega_0\rangle_2 = 2\kappa \langle W v,\omega^3\rangle_2$ shows that $s\mapsto Q_0\Gt_s\omega_0\in Y^*$ is $C^1$. With this, \eqref{eq:alpha-C1} is immediate.
\end{proof}

It is worth noting that if we let $\tilde \alpha = W\alpha$, then $\tilde\alpha$ is uniquely determined by
\begin{equation*}
    \langle \tilde\alpha(s),\omega_s\rangle_2 = 0,\quad\text{and}\quad
    Q_s(k_s + G_s)\tilde\alpha(s) = Q_sG_s\omega_s
\end{equation*}
 
\begin{prop}
\label{prop:phase}
There is $C<\infty$ such that, for all $0<\varepsilon\le\varepsilon_0$,
\begin{equation}
\label{eq:theta}
  \sup_{s\in[0,1]}|\thee(s)|\le C\varepsilon .
\end{equation}
\end{prop}
 
\begin{proof}
Applying $Q_0$ to the real part of the GP equation~\eqref{eq:real} annihilates the term
$-\varepsilon\langle\nu,\eta\rangle_2\omega_0$, and leaves, in $Y^*$,
\begin{equation}
\label{eq:Qproj}
  -\varepsilon(\et)'=L_{\perp,s}\xie(s)+q_0(s),
\end{equation}
where 
\begin{equation*}
    q_0(s)=(\re(s)-1)Q_0\Gt_s\omega_0+Q_0\operatorname{Re}\Rt(s)+\be(s)\xie(s),
\end{equation*}
and $\|q_0(s)\|_2\le C\varepsilon^2$ by \eqref{eq:rbounds}, \eqref{eq:Rsmall},
\eqref{eq:xieta} and Lemma~\ref{lem:small beta}.  By symmetry of $\ell_s$ and the
definition \eqref{eq:alpha},
\begin{align*}
  \langle\omega_0,\Gt_s\xie(s)\rangle_2
  &=\langle \Gt_s\omega_0,\xie(s)\rangle_2
  =\ell_s(\alpha(s),\xie(s))
  =\langle\alpha(s),L_{\perp,s}\xie(s)\rangle_{\mathcal{Q}-\mathcal{Q}^*} \\
  &=-\varepsilon\langle \alpha(s),(\et)'\rangle_2-\langle \alpha(s),q_0(s)\rangle_2
\end{align*}
by~\eqref{eq:Qproj}. Substituting back into \eqref{eq:pair-real} and dividing by $\varepsilon\re$ yields
\begin{equation}
\label{eq:thetaprime}
  (\thee)'(s)=\frac{\langle\alpha(s),(\et)'(s)\rangle_2}{\re(s)}+\frac{\langle\alpha(s),q_0(s)\rangle_2-\lambda(s)}{\varepsilon \re(s)},
\end{equation}
and the second term is $\caO(\varepsilon)$ by \eqref{eq:lambda} and Lemma~\ref{lem:corrector}.
 
The first term carries a factor $(\et)'$, which is not small, but it is a total
derivative up to $\caO(\varepsilon)$.  Indeed $\alpha\in C^1([0,1];Y)$,
$\et\in C([0,1];Y)$ with $(\et)'\in Y^*$, and $r$ being absolutely continuous and
bounded away from zero together justify the application of the product rule in the real triple
$Y\subset Q_0\caH_\R\subset Y^*$, namely (dropping the explicit time dependence in the notation)
\begin{equation*}
  \int_0^s\frac{\langle\alpha,(\et)'\rangle_2}{\re}d\tau
  =\left.\frac{\langle\alpha,\et\rangle_2}{\re}\right\vert_0^s
   -\int_0^s\frac{\langle\alpha',\et\rangle_2}{\re}d\tau
   +\int_0^s\frac{(\re)'}{(\re)^2}\langle\alpha,\et\rangle_2 d\tau .
\end{equation*}
The boundary term and the first integral are $\caO(\varepsilon)$ by~\eqref{eq:xieta} and Lemma \ref{lem:corrector}, while the second integral is $O(\varepsilon^2)$ with~\eqref{eq:r'}. \eqref{eq:theta} now follows by integrating \eqref{eq:thetaprime} and using $\thee(0)=0$.
\end{proof}

\subsection{Conclusion of the proof of Theorem~\ref{thm:abstract}}
 
By the transport identity and \eqref{eq:polar},\eqref{eq:decomp of chi},
\begin{align*}
    \ph(s)-\omega_s
    &=W(s)\bigl(\ze(s)-\omega_0\bigr) \\
    &= W(s)\left(\big(\mathrm{e}^{\iu\thee(s)}-1\big)\omega_0 +\mathrm{e}^{\iu\theta(s)}\big((\re(s)-1)\omega_0+\xie(s)+\mathrm{i}\et(s)\big)\right).
\end{align*}
The operators $W(s)$ are uniformly bounded on $\caQ$, and
$\Vert (\mathrm{e}^{\iu \thee}-1)\omega_0\Vert_\caQ \leq |\thee|\Vert \omega_0\Vert_\caQ$.
Hence, by Proposition \ref{prop:phase}, \eqref{eq:rbounds} and \eqref{eq:xieta},
\begin{equation*}
    \Vert\ph(s)-\omega_s\Vert_\caQ
    \leq C\left(|\thee(s)|+|\re(s)-1|+\|\xie(s)\|_\caQ+\|\et(s)\|_\caQ\right)\le D\varepsilon
\end{equation*}
uniformly on $[0,1]$. This is~\eqref{eq:Main} upon recalling that $\ph(s)=\mathrm{e}^{\frac{\iu}{\varepsilon}\int_0^s\mu_\tau d\tau}\ps(s)$, thereby concluding the proof of Theorem~\ref{thm:abstract}. \hfill$\square$

\section{Proof of Theorem \ref{thm:conditional}} \label{sect:conditional}

This is just a concrete Schr\"odinger setting for Theorem~\ref{thm:abstract}. What must be proved is the following.

\begin{prop} \label{prop: Assumptions}
    If Assumptions~\ref{as: confining potential}, \ref{as: form} hold, then with the choice 
    \begin{equation} \label{as Schrödinger}
    a_s(u,v)=\int_{\mathbb{R}^3} \left( \overline{\nabla u} \cdot \nabla v + V_s \overline{u} v\right),
\end{equation}
and
\begin{equation} \label{Q Schrödinger}
    \mathcal{Q}= \mathcal{Q}_W = \left\{ f\in H^1(\mathbb{R}^3) \ : \ \Vert f \Vert_{\mathcal{Q}_W}^2 = \int \left[ \vert \nabla f(x)\vert^2+  W(x) \vert f(x) \vert^2 \right] dx < \infty \right\},
\end{equation}
the Assumptions \ref{as: Q}--\ref{as: energy} are satisfied and $\mathcal{Q}\subseteq H^1(\mathbb{R}^3)$ is a dense subspace which is invariant under complex conjugation. 
\end{prop}

\begin{proof}[Proof of Theorem \ref{thm:conditional}]
    By Proposition~\ref{prop: Assumptions}, all assumptions of Theorem~\ref{thm:abstract} are satisfied and the claim immediately follows.
\end{proof}

We first record a structural embedding result.
\begin{lemma}
\label{lem:compact}
The embedding ${\mathcal{Q}_W}\hookrightarrow L^2(\R^3)$ is compact.  Consequently,
\begin{enumerate}
\item $-\Delta+W$ has compact resolvent and 
\item if $f_n\rightharpoonup f$ weakly in ${\mathcal{Q}_W}$, then
  $f_n\to f$ strongly in $L^p(\R^3)$ for every $p\in[2,6)$.
\end{enumerate}
\end{lemma}

\begin{proof}
Let $(f_n)_n$ be bounded in ${\mathcal{Q}_W}$, say $\|f_n\|_{{\mathcal{Q}_W}}^2\le M$. Then $\|f_n\|_{H^1}^2\le\|f_n\|_{{\mathcal{Q}_W}}^2\le M$ since $W\ge1$. Now, for every $R>0$,
\begin{equation}
\label{eq:tails}
  \int_{|x|>R}|f_n|^2
  \leq \Bigl(\einf_{|x|>R}W\Bigr)^{-1}\int_{|x|>R}W|f_n|^2
  \leq\frac{M}{\einf_{|x|>R}W} = \delta_R ,
\end{equation}
and $\delta_R\to0$ as $R\to\infty$ by Assumption~\ref{as: confining potential}, uniformly in $n$.
 
By Rellich--Kondrachov, the embedding $H^1(B_R)\hookrightarrow
L^2(B_R)$ is compact for each $R\in\mathbb N$; hence there is a
subsequence $(f_{n_k})_k$ converging in $L^2(B_R)$ for every $R$.  For such a
subsequence, \eqref{eq:tails} gives
\[
  \limsup_{k,l\to\infty}\|f_{n_k}-f_{n_l}\|_{L^2(\R^3)}^2
  \le\limsup_{k,l\to\infty}\|f_{n_k}-f_{n_l}\|_{L^2(B_R)}^2+4\delta_R
  =4\delta_R
\]
for every $R$, hence $(f_{n_k})_k$ is Cauchy in $L^2(\R^3)$.  This proves
compactness.
 
\noindent \emph{1.}  Let $A=-\Delta+W\ge1$.  Then ${\mathcal{Q}_W}$ is the domain of $A^{1/2}$ and $\|u\|_{{\mathcal{Q}_W}}=\|A^{1/2}u\|_2$, so $A^{-1/2}:L^2\to{\mathcal{Q}_W}$ is unitary and the
embedding $\iota:{\mathcal{Q}_W} \hookrightarrow L^2$ satisfies $\iota\circ A^{-1/2}=A^{-1/2}$ as a map
$L^2\to L^2$.  Hence $\iota$ is compact if and only if $A^{-1/2}$ is compact on $L^2$, which is the case if and only if $A^{-1}$ is.
 
\noindent \emph{2.}  First, $p=2$.  Since $\iota$ is bounded, $f_n\rightharpoonup f$ in
$L^2$.  If $\|f_n-f\|_2\not\to0$, there are $\delta>0$ and a subsequence with
$\|f_{n_k}-f\|_2\ge\delta$; but $(f_n)_n$ is bounded in ${\mathcal{Q}_W}$, being weakly
convergent, so compactness yields a further subsequence converging strongly in
$L^2$ to some $g$, and uniqueness of weak limits forces $g=f$, a contradiction.
For $p\in(2,6)$ we use H\"older with $\tfrac1p=\tfrac\theta2+\tfrac{1-\theta}{6}$ with
$\theta\in(0,1)$ to conclude that
\[
  \|f_n-f\|_p\le\|f_n-f\|_2^{\theta}\,\|f_n-f\|_6^{1-\theta}.
\]
The $L^p$ convergence follows from the $L^2$ convergence and the fact that $\|f_n-f\|_6$ is bounded because $(f_n)_n$ is bounded in $H^1$.
\end{proof}

We now turn to the proof of Proposition \ref{prop: Assumptions}. First of all, it is easy to see that $\mathcal{Q}_W\subseteq H^1(\mathbb{R}^3)$ is a dense subspace which is invariant under complex conjugation. Since $W\geq 1$, $\Vert u\Vert_{H^1}^2 = \int(\vert \nabla u\vert^2 + \vert u\vert^2)\leq \Vert u\Vert_{\caQ_W}^2$ so Assumption~\ref{as: Q} holds with $c_\caQ = 1$.

With $c<1$ as in Assumption \ref{as: form},
\begin{align*}
    c\Vert u \Vert_{\mathcal{Q}_W}^2 \leq \int_{\mathbb{R}^3} \left(c \vert \nabla u\vert^2 + (V_s+M) \vert u \vert^2 \right)
    \leq a_s[u] + M\Vert u\Vert_2^2.
\end{align*}
A lower bound follows similalry with $C>1$ from Assumption \ref{as: form}. This proves Assumption~\ref{as: quadratic forms}.

Before proving Assumptions~\ref{as: ground states} and \ref{as: spectral gap}, we turn to the forms $a_s,b_s$, proving first that $b_s$, see~\eqref{eq:b}, is non-negative with a unique positive element in its kernel.
\begin{lemma}
\label{lem:pf}
Let $s\in[0,1]$ and suppose that $\omega\in\mathcal{Q}_W$ with $\omega\ge0$ and
$\|\omega\|_2=1$, and that $\mu\in\R$, satisfy
\begin{equation}
\label{eq:ELfixed}
  a_s(u,\omega)+\kappa\langle u,\omega^3\rangle_2=\mu\langle u,\omega\rangle_2
  \qquad\text{for all }u\in\mathcal{Q}_W .
\end{equation}
Then,
\begin{enumerate}
\item $\mu=\min\sigma(h_s+\kappa\omega^2)$ is a simple eigenvalue and $\omega>0$ almost everywhere;
\item $b_s[u]\ge0$ for all $u\in\mathcal{Q}_W$, and $b_s[u]=0$ if and only if $u\in\C\omega$.
\end{enumerate}
\end{lemma}
 
\begin{proof}
Let $U=V_s+\kappa\omega^2$. Since $\omega\in\mathcal{Q}_W\hookrightarrow L^4$, see~\eqref{eq:interpolation}, we have that
\[
  \|\omega u\|_2^2=\langle u,\omega^2u\rangle_2\le\|\omega\|_4^2\|u\|_4^2
  \leq C  \|\omega\|_{\mathcal{Q}_W}^2\|u\|_{\mathcal{Q}_W}^2.
\]
Together with Assumption~\ref{as: quadratic forms} this gives
\begin{equation}
\label{eq:formeq}
  c_a\|u\|_{\mathcal{Q}_W}^2\ \le\ a_s[u]+\kappa\|\omega u\|_2^2+m\|u\|_2^2
  \ \le\ \bigl(C_a+\kappa C\|\omega\|_{\mathcal{Q}_W}^2\bigr)\|u\|_{\mathcal{Q}_W}^2.
\end{equation}
Hence the form of $-\Delta+U+m$ is
closed, bounded below, with form domain exactly $\mathcal{Q}_W$ and form norm equivalent
to $\|\cdot\|_{\mathcal{Q}_W}$. Because of this equivalence, we conclude that $(-\Delta+U+m)^{-1}$ is compact as in Lemma~\ref{lem:compact}(1). In particular, $-\Delta+U$ has purely discrete spectrum.

By~\eqref{eq: form1}, $V_s\ge cW-M\ge c-M$ since $W\geq1$, so $U$ is bounded below. Moreover, $U\in L^1_{\mathrm{loc}}$ because $V_s\in L^\infty_{\mathrm{loc}}$, while $\omega^2\in L^3$. By~\cite[Theorem~1]{FarisSimon}, the ground state energy $\lambda_0 = \min\sigma(-\Delta + U)$ is non-degenerate. Moreover, the proof of that theorem establishes that $\mathrm e^{-t(-\Delta+U)}$, is positivity improving.  Furthermore, the Perron--Frobenius theorem for positivity improving semigroups \cite[Thm.~XIII.44]{ReedSimonIV} shows that $\lambda_0$ is simple with an eigenfunction $\varphi>0$ almost everywhere.
 
Now, by~\eqref{eq:ELfixed}, $\omega$ is an eigenfunction of $-\Delta+U$ with eigenvalue $\mu$.  If
$\mu\neq\lambda_0$ then $\langle \varphi, \omega\rangle_2=0$, which is impossible if both $\varphi>0$ a.e and $\omega$ is non-negative. Hence $\mu=\lambda_0$, and simplicity gives
$\omega\in\C\varphi$, so $\omega>0$ a.e. This proves the first part. The second part is now immediate with $b_s[u]=\langle u,(-\Delta+U-\mu)u\rangle_{\mathcal{Q}-\mathcal{Q}^*}\ge0$.
\end{proof}

We now establish continuity of $\omega$ and $\mu$ we consider the ground state energy
\begin{align*}
    e_s = \mathcal{E}_s(\omega_s)=\inf_{f\in \mathcal{Q}_W:\Vert f \Vert_2=1} \int_{\mathbb{R}^3} \left( \vert \nabla f \vert^2 + V_s \vert f \vert^2+ \frac{\kappa}{2} \vert u \vert^4 \right).
\end{align*}
With this,
\begin{align*}
    e_s \leq \mathcal{E}_s(\omega_t) \leq e_t + \Vert V_s - V_t \Vert_{\mathcal{M}_W} \sup_{\tau\in [0,1]} \Vert \omega_\tau\Vert^2_{\mathcal{Q}_W},
\end{align*}
so that
\begin{equation} \label{eq: es cont}
    \vert e_s - e_t\vert, \vert e_s - \mathcal{E}_s(\omega_t)\vert 
    \leq \Vert V_s - V_t \Vert_{\mathcal{M}_W} \sup_{\tau\in [0,1]} \Vert \omega_\tau\Vert_{\mathcal{Q}_W}^2.
\end{equation}
Since $s\mapsto V_s$ is assumed to be continuous in $\mathcal{M}_W$, we conclude that the ground state energy $s\mapsto e_s$ is continuous and that $\lim_{t\rightarrow s} \mathcal{E}_s(\omega_t)=e_s$. 

Pick $t_n\to s\in[0,1]$.  As
$(\omega_{t_n})_n$ is uniformly bounded in $\mathcal{Q}_W$, there is a $\mathcal{Q}_W$-weakly
convergent subsequence, which we again denote by $(\omega_{t_n})_n$; let
$u\in\mathcal{Q}_W$ be its weak limit.  By Lemma \ref{lem:compact}(ii) the convergence is
strong in $L^2$ and in $L^4$, with the same limit $u$, so $\|u\|_2=1$.  Moreover
$u\ge0$ since the cone $\{v\in L^2:v\ge0\}$ is convex and closed, hence weakly
closed, and $\omega_{t_n}\ge0$.
 
Using~\eqref{eq: es cont} and the weak lower semicontinuity of $\mathcal{E}_s$, we get
\[
\mathcal{E}_s(u)
\leq\lim_{n\to\infty}\mathcal{E}_s(\omega_{t_n})=\mathcal{E}_s(\omega_s).
\]
Since $\|u\|_2=1$ and $\mathcal{E}_s(\omega_s)=e_s$ is the infimum, $\mathcal{E}_s(u)=e_s$.  As
$\omega_s$ is the unique nonnegative minimiser of $\mathcal{E}_s$, we conclude that $u=\omega_s$, namely $\omega_{t_n}\rightharpoonup\omega_2$ weakly in $\mathcal{Q}_W$.
 
It remains to upgrade this to strong convergence in $\mathcal{Q}_W$. Recalling~\eqref{eq:Nonlinear Energy} and by the
convergence of $(\omega_{t_n})_n$ in $L^4$,
\begin{equation}
\label{eq:aconv}
  a_s[\omega_{t_n}]=\mathcal{E}_s(\omega_{t_n})-\frac\kappa2\|\omega_{t_n}\|_4^4
  \;\longrightarrow\;\mathcal{E}_s(\omega_s)-\frac\kappa2\|\omega_s\|_4^4=a_s[\omega_s].
\end{equation}
By Assumption~\ref{as: quadratic forms} the quadratic form $a_s[v]+m\|v\|_2^2$ is, for this fixed $s$, an equivalent Hilbert norm on $\mathcal{Q}_W$.  Since $\|\omega_{t_n}\|_2=\|\omega_s\|_2=1$, \eqref{eq:aconv} gives $a_s[\omega_{t_n}]+m\;\longrightarrow\;a_s[\omega_s]+m$. Weak convergence and convergence of the norms implies strong convergence, hence also in $\mathcal{Q}_W$. Thus, $s\mapsto\omega_s$ is continuous in $\mathcal{Q}_W$. Continuity of $s\mapsto \mu_s$ follows from this and the identity $\mu_s = e_s + \frac{\kappa}{2} \Vert \omega_s\Vert_4^4$.

With these preliminaries, we prove Assumptions~\ref{as: spectral gap}. Note first that by Lemma \ref{lem:pf} we may write, for any $s$ and any
$u\in{\mathcal{Q}_{\mathbb{R}}}$, $u=P_su+Q_su$ and
\begin{equation}
\label{eq:bQ}
  b_s[u]=b_s[Q_su],
\end{equation}
since $b_s(\omega_s,\cdot)=0$ by \eqref{eq:ELfixed} kills both $b_s[P_su]$ and
the two cross terms. Assume by contradiction that no $g>0$ as in~\eqref{eq:b gap} exists.  Then for
every $n\in\mathbb N$ there are $s_n\in[0,1]$ and $u_n\in{\mathcal{Q}_{\mathbb{R}}}$ with
$b_{s_n}[u_n]<\tfrac1n\|Q_{s_n}u_n\|_2^2$.  Here $Q_{s_n}u_n\neq0$: otherwise
the right-hand side vanishes and $b_{s_n}[u_n]<0$, contradicting Lemma
\ref{lem:pf}(ii).  Hence
$f_n:=Q_{s_n}u_n/\|Q_{s_n}u_n\|_2$ is well defined and
\begin{equation}
\label{eq:fn}
  \langle f_n,\omega_{s_n}\rangle_2=0,
  \qquad
  \|f_n\|_2=1,
  \qquad
  0\le b_{s_n}[f_n]\le\tfrac1n.
\end{equation}
 By Assumption~\ref{as: quadratic forms},
\[
  c_a\|f_n\|_{\mathcal{Q}_W}^2\ \le\ a_{s_n}[f_n]+m\|f_n\|_2^2
  \ =\ b_{s_n}[f_n]-\kappa\|\omega_{s_n}f_n\|_2^2+(\mu_{s_n}+m)\|f_n\|_2^2
  \ \le\ \tfrac1n+M+m ,
\]
where $M = \sup_s\mu_s$, so $(f_n)_{n}$ is bounded in $\mathcal{Q}_W$. Passing to a subsequence, we may assume that it converges weakly in $\mathcal{Q}_W$ to some $f\in\mathcal{Q}_W$ and that
$s_n\to s\in[0,1]$.  By the compactness Lemma~\ref{lem:compact}, the convergence is strong in $L^2$ and
in $L^4$.  In particular $\|f\|_2=1$; and since
$\omega_{s_n}\to\omega_s$ in $L^2$ by the continuity just proved,
\eqref{eq:fn} gives
\begin{equation}
\label{eq:forth}
  \langle f ,\omega_s \rangle_2=0 .
\end{equation}
 
Now, $(a_{s_n}-a_s)[u] \leq |\int(V_{s_n}-V_s)|u|^2|\le\|V_{s_n}-V_s\|_{\MW}\|u\|_{\mathcal{Q}_W}^2$ and 
\[
  \|\omega_{s_n}^2-\omega_s^2\|_2
  \le\|\omega_{s_n}-\omega_s\|_4\bigl(\|\omega_{s_n}\|_4+\|\omega_s\|_4\bigr)
  \le C \|\omega_{s_n}-\omega_s\|_4,
\]
imply that for every $u\in\mathcal{Q}_W$,
\[
  \bigl|b_{s_n}[u]-b_s[u]\bigr|
  \le C \Bigl(\|V_{s_n}-V_s\|_{\MW}
    + \|\omega_{s_n}^2-\omega_s^2\|_2
    + |\mu_{s_n}-\mu_s|\Bigr)\|u\|_{\mathcal{Q}_W}^2
  =c_n\|u\|_{\mathcal{Q}_W}^2,
\]
where we used Assumption~\ref{as: Q} for the $\mu$-term. By Assumption~\ref{as: form}, by the continuity of $s\mapsto\omega_s$ and the continuity of $\mu$, we conclude that $c_n\to0$.  Since $\sup_n\|f_n\|_{\mathcal{Q}_W}<\infty$, \eqref{eq:fn} yields
\begin{equation}
\label{eq:bsfn}
  \left \vert b_s[f_n]-b_{s_n}[f_n]\right \vert \longrightarrow0 .
\end{equation}
It remains to take the limit of $f_n$. the form $\tilde a_s[u]:=a_s[u]+m\|u\|_2^2$ defines a norm equivalent to
$\|\cdot\|_{\mathcal{Q}_W}$; in particular $\tilde a_s$ is weakly lower semicontinuous on $\mathcal{Q}_W$. We decompose
\[
  b_s[f_n]=\tilde a_s[f_n]+\kappa\|\omega_sf_n\|_2^2-(\mu_s+m)\|f_n\|_2^2,
\]
and the last two terms converge by the $L^4$, respectively $L^2$ convergence of $(f_n)$.  Hence, the convergence~\eqref{eq:fn} $b_{s_n}[f_n]\to0$, \eqref{eq:bsfn} and the weak lower semicontinuity of $\tilde a_s$ yield
\[
  0=\lim_{n\to\infty}b_s[f_n]
  \geq \tilde a_s[f]+\kappa\|\omega_sf\|_2^2-(\mu_s+m)\|f\|_2^2
  \ =\ b_s[f]\ \ge\ 0 ,
\]
the last inequality by Lemma \ref{lem:pf}(ii).  Therefore $b_s[f]=0$, so
$f\in\C\omega_s$ by Lemma \ref{lem:pf}(ii), which contradicts~\eqref{eq:forth} since $f$ is normalized. This concludes the proof of Assumptions~\ref{as: spectral gap}.

In order to prove Assumptions~\ref{as: ground states}, it remains to establish the regularity of the ground state, namely $\omega\in C^2([0,1], \mathcal{Q}_W)$ and $\mu\in C^1([0,1], \mathbb{R})$, having already established continuity. This will be obtained by leveraging the assumed regularity of $s\mapsto V(\cdot,s)$ by through an implicit function theorem argument.

Summarizing, the minimiser $\omega_s$ satisfies the Euler--Lagrange equation
\begin{equation}
\label{eq:EL}
  a_s(u,\omega_s)+\kappa\langle u, \omega_s^3\rangle_2=\mu_s\langle u, \omega_s \rangle_2
  \qquad\text{for all }u\in\mathcal{Q}_W ,
\end{equation}
or equivalently
\begin{equation}
\label{eq:bortho}
  b_s(u,\omega_s)=0\qquad\text{for all }u\in\mathcal{Q}_W .
\end{equation}
The gap estimate can be rephrased as a coercivity condition on the form $b_s$. Indeed, if $u\in{\mathcal{Q}_{\mathbb{R}}}$ with $Q_su=u$, then
\[
  c_a\|u\|_{\mathcal{Q}_{\mathbb{R}}}^2\ \le\ a_s[u]+m\|u\|_2^2
  \ =\ b_s[u]-\kappa\|\omega_su\|_2^2+(\mu_s+m)\|u\|_2^2
  \ \le\ b_s[u]+C\|u\|_2^2 .
\]
Assumption~\ref{as: spectral gap} reads $\|u\|_2^2=\|Q_su\|_2^2\le g^{-1}b_s[u]$, namely there is $c_g>0$ such that 
\begin{equation}
\label{eq:coerc}
  b_s[u]\ \ge\ c_g\|u\|_{\mathcal{Q}_{\mathbb{R}}}^2,
\end{equation}
for all $s\in[0,1]$ and all $u\in{\mathcal{Q}_{\mathbb{R}}}$ with $\langle \omega_s, u\rangle_2=0$.
  
We now fix an open interval $I\supseteq[0,1]$ on which $s\mapsto V(\cdot,s)$
is defined and $C^2$ into $\MW$ and define
\[
  j:L^2(\R^3,\R)\to\mathcal{Q}_{\mathbb{R}}^*,\quad (jf)(v)=\langle v, f\rangle_2,
  \qquad
  A_s:\mathcal{Q}_{\mathbb{R}}\to\mathcal{Q}_{\mathbb{R}}^*,\quad (A_su)(v)=a_s(v,u),
\]
both bounded and linear, and let
\[
  T:I\times\mathcal{Q}_{\mathbb{R}}\times\R\to\mathcal{Q}_{\mathbb{R}}^*\times\R,
  \qquad
  T(s,u,\nu)=\bigl(A_su+\kappa j(u^3)-\nu j(u),\ \|u\|_2^2-1\bigr),
\]
so that\eqref{eq:GS of a} reads
\begin{equation*}
    T(s,\omega_s,\mu_s)=(0,0).
\end{equation*}
 
For $V\in\MW$ and $u,v\in\mathcal{Q}_W$, Cauchy--Schwarz with weight
$W$ gives
\[ \Bigl|\int_{\R^3}V\,uv\Bigr|\le\|V\|_{\MW}\int_{\R^3}W|u||v|\le\|V\|_{\MW}\|u\|_{\mathcal{Q}_W}\|v\|_{\mathcal{Q}_W} ,
\]
so $V\mapsto\bigl((u,v)\mapsto\int Vuv\bigr)$ is a bounded linear map $\MW\to\mathcal B(\mathcal{Q}_{\mathbb{R}},\mathcal{Q}_{\mathbb{R}}^*)$. Composing it with the $C^2$ curve
$s\mapsto V(\cdot,s)$ of Assumption~\ref{as: form} shows that $s\mapsto A_s$
is $C^2$ into $\mathcal B(\mathcal{Q}_{\mathbb{R}},\mathcal{Q}_{\mathbb{R}}^*)$.  The map $u\mapsto\kappa j(u^3)$ is a
bounded homogeneous cubic polynomial, hence real-analytic: the trilinear map
$\mathcal{Q}_{\mathbb{R}}^3\to L^2(\R^3,\R)$, $(u,v,w)\mapsto uvw$, is bounded by H\"older with
$\tfrac12=3\cdot\tfrac16$ and $\mathcal{Q}_W\hookrightarrow L^6$, and $j$ is bounded.
Finally $u\mapsto\|u\|_2^2$ is a bounded quadratic form on $\mathcal{Q}_{\mathbb{R}}$ by Assumption~\ref{as: Q}.
Altogether, this shows that $T$ is $C^2$ on $I\times\mathcal{Q}_{\mathbb{R}}\times\R$.
 
Its partial Jacobian in the last two variables is
\[
  \bigl(D_{(u,\nu)}T\bigr)(s,f,\tau)[v,\eta]
  =\bigl(A_sv+3\kappa j(f^2v)-\tau j(v)-\eta\,j(f),\ 2\langle f, v\rangle_2\bigr).
\]
The following result is key for the applicability of the implicit function theorem: The Jacobian is an isomorphism at the ground state
 
\begin{prop}
\label{prop:iso}
For every $s\in[0,1]$ the map
$\bigl(D_{(u,\nu)}T\bigr)(s,\omega_s,\mu_s):\mathcal{Q}_{\mathbb{R}}\times\R\to\mathcal{Q}_{\mathbb{R}}^*\times\R$
is a bounded linear isomorphism.
\end{prop}
 
\begin{proof}
Given $(F,r)\in\mathcal{Q}_{\mathbb{R}}^*\times\R$ we must produce a unique
$(v,\eta)\in\mathcal{Q}_{\mathbb{R}}\times\R$ with
\begin{equation}
\label{eq:jac}
  \begin{cases}
    F(z)=\ell(z,v)-\eta\langle z, \omega_s\rangle_2 &\text{for all }z\in\mathcal{Q}_{\mathbb{R}},\\[2pt]
    r=2\langle \omega_s, v\rangle_2,
  \end{cases}
\end{equation}
where
\begin{align}
  \ell(z,v)&=(A_sv)(z)+3\kappa j(\omega_s^2v)(z)-\mu_s j(v)(z) \nonumber \\
  &= b_s(z,v)+2\kappa\langle z, \omega_s^2v\rangle_2 \label{eq:elldef}\\
  &=\bigl(a_s(z,v)+m\langle z, v\rangle_2\bigr)
  +3\kappa\langle z, \omega_s^2v\rangle_2-(\mu_s+m)\langle z, v\rangle_2. \nonumber
\end{align}
This is a symmetric and real form on $\mathcal{Q}_{\mathbb{R}}$. Using Assumption~\ref{as: quadratic forms} for the first bracket, Assumption~\ref{as: Q} for the last term, and
$|\langle z, \omega_s^2v\rangle_2|\le\|\omega_s\|_4^2\|z\|_4\|v\|_4$ with
$\mathcal{Q}_W\hookrightarrow L^4$ for the middle one, $\ell$ is bounded, namely $|\ell(z,v)|\le C_\ell\|z\|_{\mathcal{Q}_{\mathbb{R}}}\|v\|_{\mathcal{Q}_{\mathbb{R}}}$.
 
We now split $\mathcal{Q}_{\mathbb{R}}=\R\omega_s\oplus Y$ where $Y:=\{z\in\mathcal{Q}_{\mathbb{R}}:\langle \omega_s, z\rangle_2=0\}$. The second
equation of \eqref{eq:jac} fixes the first component of $v$, namely
\begin{equation}
\label{eq:vsplit}
  v=\tfrac r2\,\omega_s+y,\qquad y\in Y. 
\end{equation}
The decomposition also means that the first equation of \eqref{eq:jac} holds for all
$z\in\mathcal{Q}_{\mathbb{R}}$ if and only if it holds for all $z\in Y$ and for $z=\omega_s$. We consider the two cases separately.
 
The case $z\in Y$. In that case $\langle z,\omega_s\rangle=0$ and with the decomposition~\eqref{eq:vsplit}, the equation becomes
\begin{equation}
\label{eq:onY}
  \ell(z,y)=F(z)-\tfrac r2\,\ell(z,\omega_s)
  = F(z)-\kappa r\langle z, \omega_s^3\rangle_2.
\end{equation}
The right-hand side is a bounded linear functional on $Y$,
since $|\langle z, \omega_s^3\rangle_2|\le\|z\|_4\|\omega_s\|_4^3$.  The form $\ell$ is
coercive on $Y$: by \eqref{eq:elldef}, $\kappa>0$ and~\eqref{eq:coerc},
\[
  \ell(y,y)=b_s[y]+2\kappa\|\omega_sy\|_2^2\ \ge\ b_s[y]\ \ge\ c_g\|y\|_{\mathcal{Q}_{\mathbb{R}}}^2
  \qquad (y\in Y).
\]
By the Lax--Milgram theorem there is therefore a unique $y\in Y$ solving
\eqref{eq:onY}, and $(F,r)\mapsto y$ is bounded.
 
The case $z=\omega_s$.  Since $\omega_s$ is normalized, the first
equation of \eqref{eq:jac} reads $F(\omega_s)=\ell(\omega_s,v)-\eta$, namely
\begin{equation}
\label{eq:eta}
  \eta=\ell(\omega_s,y)+\tfrac r2\,\ell(\omega_s,\omega_s)-F(\omega_s)
  =2\kappa\langle \omega_s^3, y\rangle_2 +\kappa r\|\omega_s\|_4^4-F(\omega_s),
\end{equation}
which determines $\eta$ uniquely and depends boundedly on $(F,r)$, since $y$
does.
 
Thus \eqref{eq:jac} has exactly one solution $(v,\eta)$, and
$(F,r)\mapsto(v,\eta)$ is bounded. 
\end{proof}
 
With this, it remains to apply the implicit function theorem to conclude. Fix $s_0\in[0,1]$.  By Proposition \ref{prop:iso} the map $T$ is $C^2$ near
$(s_0,\omega_{s_0},\mu_{s_0})$ and $\bigl(D_{(u,\nu)}T\bigr)(s_0,\omega_{s_0},\mu_{s_0})$
is an isomorphism, so the implicit function theorem provides a relatively open
interval $I_0\subseteq[0,1]$ containing $s_0$, a neighbourhood
$\mathcal U\subseteq\mathcal{Q}_{\mathbb{R}}\times\R$ of $(\omega_{s_0},\mu_{s_0})$, and a map
\[
  I_0\ni s\longmapsto(\tilde u_s,\tilde\nu_s)\in\mathcal U
  \qquad\text{of class }C^2,
\]
such that $T(s,\tilde u_s,\tilde\nu_s)=(0,0)$ on $I_0$ and such that
$(\tilde u_s,\tilde\nu_s)$ is the only zero of $T(s,\cdot,\cdot)$ in
$\mathcal U$, for each $s\in I_0$. The family $s\mapsto(\omega_s,\mu_s)$ also solves $T(s,\cdot,\cdot)=(0,0)$, and
it is continuous at $s_0$. Hence
$(\omega_s,\mu_s)\in\mathcal U$ for all $s$ in a possibly smaller relatively
open interval $I_1\ni s_0$, and $(\omega_s,\mu_s)=(\tilde u_s,\tilde\nu_s)$ for all $s\in I_1$ by uniqueness. Hence $s\mapsto\omega_s$ and $s\mapsto \mu_s$ are $C^2$ on $I_1$. Since $s_0$ is arbitrary, this completes the proof of Assumption~\ref{as: ground states}.

It remains to check that Assumption~\ref{as: PDE input} supplies Assumptions~\ref{as: existence dynamics} and~\ref{as: energy}. The right-hand side of \eqref{eq:PDE weak epsilon} is $a_s(u,\psi^{(\varepsilon)}(s)) + \kappa\langle |\psi^{(\varepsilon)}(s)|^2 u,\psi^{(\varepsilon)}(s)\rangle_2$, so the solution $\psi^{(\varepsilon)}$ of Assumption~\ref{as: PDE input} solves~\eqref{eq:Form epsilon GP} with the same regularity, which is Assumption~\ref{as: existence dynamics}. For Assumption~\ref{as: energy}, the weighted Cauchy--Schwarz estimate used before shows that $\Phi:\mathcal M_W\to B(\mathcal Q_W,\mathcal Q_W^*)$,
$\Phi(f)(u,v)=\int_{\mathbb R^3} f\,\overline{u}\,v\,dx$, is bounded and linear with
$\|\Phi\|\le 1$. Since $a_s = a^{(0)} + \Phi(V(\cdot,s))$, where
$a^{(0)}(u,v)=\int_{\mathbb R^3}\nabla\overline{u}\cdot\nabla v\,dx$ does not depend on $s$, and since $s\mapsto V(\cdot,s)$ is $C^2$ into $\mathcal M_W$ by Assumption~\ref{as: form}, we get $a\in C^2([0,1];B(\mathcal Q_W,\mathcal Q_W^*))$ with $a'_s=\Phi(\partial_s V(\cdot,s))$,
that is,
\[
  a'_s[u] = \int_{\mathbb R^3} (\partial_s V)(x,s)\,|u(x)|^2\,dx ,
  \qquad u\in\mathcal Q_W ,
\]
where $\partial_s V$ denotes the derivative in $\mathcal M_W$. Moreover, for
$u\in\mathcal Q_W$ the functional $E_s(u)$ of \eqref{eq:Nonlinear Energy} is exactly the integral in~\eqref{As3GP}. All of its terms are finite, because
$|\int V(\cdot,s)|u|^2|\le\|V(\cdot,s)\|_{\mathcal M_W}\|u\|_{\mathcal Q_W}^2$ and
$\mathcal Q_W\subset L^4$. Hence the energy $E_s$ of Assumption~\ref{as: PDE input} equals $\mathcal{E}_s(\psi^{(\varepsilon)}(s))$. Therefore $s\mapsto \mathcal{E}_s(\psi^{(\varepsilon)}(s))$ is
absolutely continuous, and for almost every $s$ the balance law \eqref{eq: balance law} reads
\[
  \frac{d}{ds}\mathcal{E}_s(\psi^{(\varepsilon)}(s))
  = \int_{\mathbb R^3}(\partial_s V)(x,s)\,|\psi^{(\varepsilon)}_s(x)|^2\,dx
  = a'_s[\psi^{(\varepsilon)}(s)] ,
\]
which is \eqref{as: balance law}.

This concludes the proof of Proposition~\ref{prop: Assumptions}.

\section{Proof of Theorem \ref{thm:subquadratic}} \label{sect:subquadratic}

By Theorem \ref{thm:conditional} we just need to check that GP equation admits a global solution and that the balance law \eqref{as: balance law} holds. For the subquadratic case this is known to hold for all times, see \cite[Theorem 1.7]{carles2009nonlinear}.

\section{Proof of Theorem \ref{thm:poly growth}} \label{sect:polygrowth}

We will use Theorem \ref{thm:conditional} to prove Theorem \ref{thm:poly growth}. The main technical work is to establish the existence of the solution and to verify that the balance law holds. We will prove an adaptation of~\cite{carles2015sharp} to the time-dependent setting. It is a rather standard contraction argument via Duhamel's formula and Strichartz estimates~\cite{YajimaZhang2004, Mizutani2014}. 

We first check that Assumptions \ref{as: confining potential}, \ref{as: form} are satisfied. In the following we will always assume that we have picked a Lipschitz continuous representative of $r_s$. As $r\in C^2([0,1], W^{1,\infty}(\mathbb{R}^3, \mathbb{R}))$ we get that $\vert r_s(x) \vert$ is uniformly bounded in $s$ and $x$, which readily implies \eqref{eq: form1} and Assumption~\ref{as: confining potential}. The rest of Assumption~\ref{as: form} follows easily from the fact that $\Vert f\Vert_{\mathcal{M}_W} \leq \Vert f \Vert_\infty$ as $W\geq 1$.

We are left to verify Assumption \ref{as: PDE input}, i.e. show existence of a weak solution of \eqref{eq:PDE weak epsilon} in the space $C([0,1], \mathcal{Q})$ and the balance law. 
The proof proceeds in three steps. First, we prove the necessary estimates to obtain local existence via a standard contraction mapping argument and the Duhamel formula. Next we will control the energy at the given regularity and use it to bootstrap the local solutions to global ones.

To simplify the notation, we will pass to physical time $t$ for the rest of the proof. We set
\begin{align*}
    B_t=r_{\varepsilon t}, \qquad \mathcal{M}(u)=\vert u \vert^2 u, \qquad S(t)= e^{-it(-\Delta+W)}.
\end{align*}
In the following we will try to establish the existence of $u(t)=\ps(\varepsilon t)$ which solves
\begin{equation}
    i \partial_t u = (-\Delta+W)u + B(t)u+\kappa \mathcal{M}(u), \qquad u(t_0)=u_0
\end{equation}
in $\mathcal{Q}_W^*$.

\subsection{Local existence}

We collect the relevant estimates to make the usual contraction argument work for
\begin{equation} \label{def: Phi}
    \Phi(u)(t)= S(t-t_0)u_0-i \int_{t_0}^t S(t-\tau) (B_\tau u_\tau+\kappa \mathcal{M}(u_\tau)) d\tau 
\end{equation}
on $X(J)=C(J; \mathcal{Q}_W)\cap L^p(J, L^\infty(\mathbb{R}^3))$, equipped with the natural norm
\begin{equation}
    \Vert u \Vert_{X(J)}:= \Vert u \Vert_{L^\infty(J; \mathcal{Q}_W)} + \Vert u \Vert_{L^p(J; L^\infty(\mathbb{R}^3))},
\end{equation}
with $p$ given by 
\begin{equation} \label{p}
       p = \frac{2(m+1)}{m}
\end{equation}
and $J=[t_0, t_0+T]\subseteq [0, \varepsilon^{-1}]$ is a closed interval of length at most $1$. Here, $m$ is the exponent associated with the growth at infinity of the potential $W$. 

\begin{lemma}
   Let $W, r, \kappa, \varepsilon$ as in Theorem \ref{thm:poly growth}, $p$ as in \eqref{p}, and $J=[t_0, t_0+T]\subseteq [0, \varepsilon^{-1}]$ with $0<T\leq 1$. Then there exists a constant $C>0$, independent of $J$, such that for all $u\in X(J), u_0\in \mathcal{Q}_W\cap L^\infty(\mathbb{R}^3)$ we have
    \begin{equation} \label{eq: local existence}
    \begin{split}
        \Vert \Phi(u) \Vert_{X(J)} &\leq C \Vert u_0 \Vert_{\mathcal{Q}_W} + CT \Vert u \Vert_{X(J)} + CT^{\gamma} \Vert u \Vert_{X(J)}^3, \\
        \Vert \Phi(u) - \Phi(v)\Vert_{X(J)} &\leq C \left( T + T^\gamma (\Vert u\Vert_{X(J)}^2+ \Vert v\Vert_{X(J)}^2) \right)\Vert u-v\Vert_{X(J)}
    \end{split}
    \end{equation}
   where $\gamma=1-2/p>0$.
\end{lemma}
\begin{proof}
The crucial bound here is the following Strichartz estimate: For $f\in \mathcal{Q}_W$
\begin{equation} \label{eq: Strichartz}
    \Vert S(\cdot) f \Vert_{L^p([-1,1]; L^\infty)} \leq C \Vert f \Vert_{\mathcal{Q}_W}.
\end{equation}
Using \cite[Theorem 1.4, Remark 1.7]{Mizutani2014} with
    \begin{align*}
        p=\frac{2(m+1)}{m}, \qquad q=\frac{6(m+1)}{m+3}, \qquad \delta= \frac{m-2}{2(m+1)}, \qquad \sigma = \frac{m+4}{2(m+1)},
    \end{align*}
yields on time intervals of length at most one
\begin{equation}\label{eq:miz}
 \Vert S(\cdot)f\Vert_{L^p(J; L^q)}\le C\Vert (-\Delta +W)^{\delta/2}f\Vert_2.
\end{equation}
The $L^q$ elliptic norm equivalence for $0\le\sigma\le1$ gives
\begin{equation}\label{eq:elliptic}
 \Vert g \Vert_{W^{\sigma,q}}\le C\bigl(\Vert (-\Delta+W)^{\sigma/2}g\Vert_q+\Vert g\Vert_q\bigr).
\end{equation}
Using \eqref{eq:miz}, \eqref{eq:elliptic}, $\delta+\sigma=1$, and $-\Delta+W\ge1$ gives
\begin{align*}\label{eq:sigmaq}
 \Vert S(\cdot)f\Vert_{L^p(J; W^{\sigma,q})}
 &\le C\bigl(\Vert S(\cdot)(-\Delta+W)^{\sigma/2}f\Vert_{L^p(J;L^q)}
+ \Vert S(\cdot)f\Vert_{L^p(J;L^q)}\bigr)\notag\\
 &\le C\bigl(\Vert (-\Delta+W)^{(\sigma+\delta)/2}f\Vert_2+\Vert (-\Delta+W)^{\delta/2}f\Vert_2\bigr) \\
 &\le C\Vert f\Vert_{\mathcal{Q}_W}.
\end{align*}
Here powers are first applied on a dense smooth core and then extended by continuity.
The Sobolev embedding $W^{\sigma,q}(\R^3)\hookrightarrow L^\infty(\R^3)$ then proves \eqref{eq: Strichartz}. 

With \eqref{eq: Strichartz} and since $S(\cdot)$ is unitary on $\mathcal{Q}_W$, we immediately obtain that

\begin{equation} \label{eq: Strichartz 2}
    \Vert S(\cdot-t_0) f \Vert_{X(J)} \leq C \Vert f \Vert_{\mathcal{Q}_W}.
\end{equation}

Writing $|f|^2f-|g|^2g=|f|^2(f-g)+(|f|^2-|g|^2)g$ and using $\big||f|^2-|g|^2\big|,\ |f^2-g^2|\le(|f|+|g|)|f-g|$ gives
\begin{align*}
|\mathcal{M}(f)-\mathcal{M}(g)|&
\leq\frac32\big(|f|^2+|g|^2\big)|f-g|,\\
|\nabla\mathcal{M}(f)-\nabla\mathcal{M}(g)|
&\leq 3|f|^2|\nabla (f-g)|+3(|f|+|g|)|f-g||\nabla g|.
\end{align*}
Hence,
\begin{align*}
\Vert W^{1/2}(\mathcal{M}(f)-\mathcal{M}(g))\Vert_{2}&\le 3(\Vert f\Vert_{\infty}^2+\Vert g\Vert_{\infty}^2)\Vert W^{1/2}(f-g)\Vert_{2},\\
\Vert \nabla(\mathcal{M}(f)-\mathcal{M}(g))\Vert_{2}&\le 3(\Vert f\Vert_{\infty}^2+\Vert g\Vert_{\infty}^2)\Vert \nabla (f-g)\Vert_{2}
+3\big(\Vert f\Vert_{\infty}+\Vert g\Vert_{\infty}\big)\Vert\nabla g\Vert_{2}\Vert f-g\Vert_{\infty}.
\end{align*}
This yields
\begin{align*}
\Vert \mathcal{M}(f)-\mathcal{M}(g)\Vert_{\mathcal{Q}_W} 
&\leq 3 \left( \Vert f \Vert_\infty^2+\Vert g \Vert_\infty^2\right) \Vert f -g \Vert_{\mathcal{Q}_W} \\
&\quad+3(\Vert f \Vert_\infty+\Vert g \Vert_\infty)(\Vert f \Vert_{\mathcal{Q}_W} + \Vert g \Vert_{\mathcal{Q}_W}) \Vert f -g \Vert_\infty,
\end{align*}
and
\begin{equation*}
    \Vert \mathcal{M}(f)\Vert_{\mathcal{Q}_W} 
    \leq 3 \Vert f \Vert_\infty^2 \Vert f \Vert_{\mathcal{Q}_W}
\end{equation*}
by setting $g=0$. This in turn implies, using Hölder's inequality in time, that for all $u,v\in X(J)$ we have
\begin{equation} \label{eq: N}
    \begin{split}
        \Vert \mathcal{M}(u)\Vert_{L^1(J;{\mathcal{Q}_W})} &\leq C T^\gamma \Vert u \Vert_{X(J)}^3 \\
        \Vert \mathcal{M}(u)-\mathcal{M}(v)\Vert_{L^1(J; {\mathcal{Q}_W})} &\leq C T^\gamma \left( \Vert u \Vert_{X(J)}^2+\Vert v\Vert_{X(J)}^2 \right) \Vert u-v\Vert_{X(J)}.
    \end{split}
\end{equation}
Using the product rule and $\sup_{s\in [0,1]} \Vert r_s \Vert_{W^{1, \infty}}<\infty$ we get for all $f\in {\mathcal{Q}_W}\cap L^\infty(\mathbb{R}^3)$ and all $\xi\in [0,1]$
\begin{equation} \label{eq: b}
        \Vert B_\xi f \Vert_{\mathcal{Q}_W} \leq C \Vert f \Vert_{\mathcal{Q}_W}.
\end{equation}
Combining \eqref{eq: N} and \eqref{eq: b} with $S(\tau)$ being unitary on ${\mathcal{Q}_W}$ for every $\tau$, we get
\begin{align*}
        \Vert \Phi(u) \Vert_{L^\infty(J;{\mathcal{Q}_W})} &\leq \Vert u_0\Vert_{\mathcal{Q}_W} + C \Vert u \Vert_{C(J;{\mathcal{Q}_W})} + \kappa\Vert \mathcal{M}(u)\Vert_{L^1(J;{\mathcal{Q}_W})} \\
        &\leq \Vert u_0\Vert_{\mathcal{Q}_W} + CT \Vert u\Vert_{X(J)} + CT^\gamma \Vert u \Vert_{X(J)}^3.
\end{align*}
On the other hand, using Minkowski's inequality, \eqref{eq: Strichartz 2}, and \eqref{eq: N}, we get
\begin{align*}
    \Vert \Phi(u)\Vert_{L^p(J; L^\infty)}
    &\leq C\Vert u_0\Vert_{\mathcal{Q}_W} +C \Vert u\Vert_{L^1(J;{\mathcal{Q}_W})}+ C\Vert \mathcal{M}(u) \Vert_{L^1(J; {\mathcal{Q}_W})} \\
    &\leq C\Vert u_0\Vert_{\mathcal{Q}_W} + CT \Vert u \Vert_{X(J)} + CT^\gamma \Vert u \Vert_{X(J)}^3.
\end{align*}
Hence, we obtain the first inequality in \eqref{eq: local existence}. The same argument proves also the second inequality in \eqref{eq: local existence}.
\end{proof}

The estimates in \eqref{eq: local existence} tells us that for sufficiently small time $T$ depending only on $\Vert u_0\Vert_{\mathcal{Q}_W}$, the map $\Phi$ is a contraction on a sufficiently large ball in $X(J)$. The contraction mapping principle then gives us a unique fixed point of $\Phi$ on said ball. The important part is that the estimates do not depend on $t_0$, but only on $\Vert u_0\Vert_{\mathcal{Q}_W}$.

\subsection{Energy balance} 

We will show now that the energy corresponding to the local solution is absolutely continuous and the derivative satisfies the balance law.

The energy cannot simply be differentiated due to the presence of unbounded operators. We justify the balance law \eqref{eq: balance law} by first regularizing and then taking a derivative. Define
\begin{equation}\label{eq:Jn}
 J_n=e^{-(-\Delta+W)/n}.
\end{equation} 
We have that $J_n\to1 $ strongly on $L^2$ and ${\mathcal{Q}_W}$. Furthermore, we have the operator bounds
\begin{equation} \label{eq: Jn contraction}
    \Vert J_n\Vert _{L^2\to L^2}\leq 1, \qquad \Vert J_n\Vert _{{\mathcal{Q}_W}\to{\mathcal{Q}_W}}\le1.
\end{equation}
Since $W\ge0$, heat-semigroup domination gives (see \cite[Theorem A.2.7, Theorem B.1.1]{simon1982schrodinger})
\begin{equation}\label{eq:heat}
 |J_nf|\le e^{\Delta/n}|f|,\qquad
 \Vert J_nf\Vert _\infty\le\Vert f\Vert _\infty,
 \qquad \Vert J_nf\Vert _\infty\le C_n\Vert f\Vert _2.
\end{equation}
For fixed $n$, $J_n:L^2\to\mathrm{Dom}((-\Delta+W)^k)$ is bounded for every integer $k\ge1$, since $\sup_{\lambda\ge1}\lambda^ke^{-\lambda/n}<\infty$.

Consider the regularized equation
\begin{equation}\label{eq:regularized}
 i\partial_tu_n=(-\Delta+W)u_n+J_nB(t)J_n u_n
             +\kappa J_n\mathcal{M}(J_nu_n),\qquad u_n(t_0)=J_nf.
\end{equation}
The regularized perturbations $J_nB(t)J_n, J_n\mathcal{M}(J_n \cdot)$ are locally Lipschitz from $L^2$ into ${\mathcal{Q}_W}$, by \eqref{eq:heat} and $\Vert (-\Delta+W)J_n\Vert _{L^2\to L^2}<\infty$. Thus, using again the contraction mapping argument via Duhamel one gets
$u_n\in C(J;{\mathcal{Q}_W})\cap C^1(J;L^2)$. Since $J_n$ and $B$ are self-adjoint multiplication/smoothing operators the regularized equation conserves $\Vert u_n(t)\Vert _2=\Vert J_n f\Vert_2$. 

Define the regularized energy
\begin{equation}\label{eq:En}
 E_n(t,v)=\Vert v\Vert _{\mathcal{Q}_W}^2+\int_{\mathbb{R}^3}B(t)|J_nv|^2
              +\frac\kappa2\Vert J_nv\Vert _4^4.
\end{equation}
Now the classical chain rule is legitimate and we compute
\begin{equation}\label{eq:Enbalance}
 \frac{d}{d t}E_n(t,u_n(t))=\int_{\mathbb{R}^3}B'(t)|J_nu_n(t)|^2.
\end{equation}
In order to obtain a statement about absolute integrability and the derivative almost everywhere, we consider the integrated version of \eqref{eq:Enbalance}
\begin{equation} \label{eq:En integrated}
    E_n(t, u_n(t))=E_n(t_0, u_n(t_0))+ \int_{t_0}^t \left(\int_{\mathbb{R}^3}B'(\tau)|J_nu_n(\tau)|^2\right) d\tau.
\end{equation}

Now we want to take the limit $n\rightarrow \infty$ and recover the desired balance law
\begin{align*}
    E_t = E_{t_0} + \int_{t_0}^t \left( \int_{\mathbb{R}^3} B'(\tau) \vert u(\tau)\vert^2 \right) d\tau.
\end{align*}

Recall that
\begin{align*}
    E_n(t, u_n(t)) &= \Vert u_n(t) \Vert_{\mathcal{Q}_W}^2+\int_{\mathbb{R}^3} B(t) \vert J_n u_n(t) \vert^2 + \frac{\kappa}{2} \Vert J_n u_n(t)\Vert_4^4,\\
    E_{\varepsilon t} &= \Vert u(t) \Vert_{\mathcal{Q}_W}^2 + \int_{\mathbb{R}^3} B(t) \vert u(t)\Vert^2 + \frac{\kappa}{2} \Vert u(t)\Vert_4^4.
\end{align*}
Each of the individual terms converge. This is immediate from the following two estimates which we will prove in the rest of this section. We show that, after potentially shrinking $J$,
\begin{equation} \label{eq: conv regularization}
\sup_{t\in J} \Vert u_n(t)-u(t) \Vert_{\mathcal{Q}_W} \rightarrow 0,
\qquad \sup_{t\in J} \Vert J_n u_n(t)-u(t) \Vert_{\mathcal{Q}_W} \rightarrow 0.
\end{equation}

In the following we assume that we have already shrunk $J$ in such a way that 
\begin{equation} \label{eq: R}
    \sup_n \Vert u_n \Vert_{X(J)}=:R<\infty.
\end{equation}
This is possible as by construction the radius of the ball needed for the contraction mapping argument to work is the size of the initial conditions $J_n u_0$ and the size of the integrands in the Duhamel formula. However, due to \eqref{eq: Jn contraction} the choice for the unregularized problem works.

In a first step we will derive pointwise estimates in ${\mathcal{Q}_W}$. Using dominated convergence we will then upgrade them to estimates in $L^1(J;{\mathcal{Q}_W})$. The fixed point equation will then allow us to finally derive \eqref{eq: conv regularization}.

We pick $g\in {\mathcal{Q}_W}\cap L^\infty$ and write $g_n=J_n g$. Our first goal is to prove
\begin{equation} \label{eq: pointwise N}
    \mathcal{M}(g_n) \rightarrow \mathcal{M}(g) \qquad \text{in } {\mathcal{Q}_W}.
\end{equation}
Using \eqref{eq:heat} we have $\sup_n \Vert g_n \Vert_\infty \leq \Vert g \Vert_\infty$ Thus, we have $\vert \mathcal{M}(g_n)-\mathcal{M}(g)\vert \leq C \Vert g \Vert_\infty^2 \vert g_n - g\vert$ and by strong convergence
\begin{equation} \label{eq: W N}
    \Vert W^{1/2} (\mathcal{M}(g_n)-\mathcal{M}(g))\Vert_2 \leq C \Vert g \Vert_\infty^2 \Vert W^{1/2} (g_n-g)\Vert_2 \rightarrow 0.
\end{equation}
Now pick a subsequence along which we have pointwise almost everywhere $g_n \rightarrow g$. We write
\begin{align*}
    \Vert \nabla \mathcal{M}(g_n) - \nabla \mathcal{M}(g) \Vert_2 
    &\leq 2 \Vert \vert g_n\vert^2 \nabla(g_n-g) \Vert_2 + \Vert g_n^2 \nabla \left( \overline{g_n}-\overline{g} \right) \Vert_2 \\ &\qquad+ 2 \Vert (\vert g_n \vert^2-\vert g \vert^2) \nabla g\Vert_2  + \Vert(g_n^2-g^2) \nabla \overline{g} \Vert_2 \\
    &\leq 3\Vert g\Vert_\infty^2 \Vert g_n-g\Vert_{\mathcal{Q}_W}+2 \Vert (\vert g_n \vert^2-\vert g \vert^2) \nabla g\Vert_2  + \Vert(g_n^2-g^2) \nabla \overline{g} \Vert_2.
\end{align*}
The first term on the last line tends to zero due to strong convergence in ${\mathcal{Q}_W}$. The remaining two terms converge to zero by dominated convergence. Combining this with \eqref{eq: W N} we get \eqref{eq: pointwise N}. 

Now we compute
\begin{align*}
    \Vert J_n \mathcal{M}(J_n g) - \mathcal{M}(g) \Vert_{\mathcal{Q}_W}
    &\leq \Vert J_n(\mathcal{M}(J_n g) - \mathcal{M}(g))\Vert_{\mathcal{Q}_W} + \Vert (J_n-I)\mathcal{M}(g) \Vert_{\mathcal{Q}_W} \\
    &\leq \Vert \mathcal{M}(J_n g) - \mathcal{M}(g) \Vert_{\mathcal{Q}_W} + \Vert (J_n-I) \mathcal{M}(g) \Vert_{\mathcal{Q}_W} \rightarrow 0.
\end{align*}
For the convergence of the first term on the last line we have used \eqref{eq: pointwise N} and for the second term the strong convergence in ${\mathcal{Q}_W}$. Thus, by 
\begin{align*}
    \Vert J_n \mathcal{M}(J_n u(t))-\mathcal{M}(u(t)) \Vert_{\mathcal{Q}_W} \leq C \Vert u(t)\Vert_\infty^2 \Vert u(t) \Vert_{\mathcal{Q}_W}
\end{align*}
and dominated convergence we get
\begin{equation}
    J_n \mathcal{M}(J_n u) \rightarrow \mathcal{M}(u) \qquad \text{in } L^1(J; {\mathcal{Q}_W}).
\end{equation}
Similarly, one shows
\begin{equation}
    J_n B J_n u \rightarrow B u \qquad \text{in } L^1(J; {\mathcal{Q}_W}).
\end{equation}
Combining the last two estimates yields
\begin{equation} \label{eq: rn}
    R_n := J_n B J_n u + \kappa J_n \mathcal{M}(J_n u) - B u - \kappa \mathcal{M}(u) \rightarrow 0 \qquad \text{in } L^1(J; {\mathcal{Q}_W}).
\end{equation}
We are now in the position to use the equation to upgrade our estimates. Since $J_n$ does not increase either $\Vert \cdot\Vert_{\mathcal{Q}_W}$ or $\Vert \cdot\Vert_\infty$, see~\eqref{eq: Jn contraction}, the contraction constant of all $\Phi_n$ is uniform, and by possibly shrinking $J$ further, we assume that it is less or equal to $1/2$. Then we get
\begin{align*}
    \Vert u_n - u \Vert_{X(J)} 
    &= \Vert \Phi_n(u_n)-\Phi(u)\Vert_{X(J)} \\
    &\leq \Vert \Phi_n(u_n)-\Phi_n(u)\Vert_{X(J)} + \Vert \Phi_n(u) - \Phi(u) \Vert_{X(J)} \\
    &\leq \frac{1}{2} \Vert u_n - u \Vert_{X(J)} + C \Vert (J_n-I)u_0\Vert_{\mathcal{Q}_W} + C \Vert R_n \Vert_{L^1(J; {\mathcal{Q}_W})},
\end{align*}
where we have used \eqref{eq: Strichartz 2} to pass to the last line. Thus, using \eqref{eq: rn} we get
\begin{equation} \label{eq: Linfty}
    \sup_{t\in J} \Vert u_n(t)-u(t) \Vert_{\mathcal{Q}_W} \rightarrow 0.
\end{equation}
Furthermore, using \eqref{eq: Jn contraction}, \eqref{eq: Linfty} and strong convergence in ${\mathcal{Q}_W}$ we get
\begin{align*}
    \Vert J_n u_n(t) - u(t) \Vert_{L^\infty(J; {\mathcal{Q}_W})}
    &\leq \Vert J_n(u_n-u)\Vert_{L^\infty(J; {\mathcal{Q}_W})} + \Vert (J_n-I)u\Vert_{L^\infty(J; {\mathcal{Q}_W})} \\
    &\leq \Vert u_n -u\Vert_{L^\infty(J; {\mathcal{Q}_W})} + \Vert (J_n-I)u \Vert_{L^\infty(J; {\mathcal{Q}_W})} \rightarrow 0.
\end{align*}
Which yields the second inequality in \eqref{eq: conv regularization}.

\subsection{Global existence} 
Assume now that the weak solution $u$ exists on the interval $[0, t]$. Lemma \ref{lem:mass} yields that $\Vert u(\tau,\cdot)\Vert_2=1$ for all $\tau \in [0,t]$. Thus, by the balance law~\eqref{eq: balance law},
\begin{align*}
    \vert E_{\varepsilon t}-E_0 \vert \leq \int_{0}^t \Vert \partial_\tau B_\tau\Vert_\infty d\tau
    \leq \int_0^1 \Vert \partial_s r_s \Vert_\infty ds.
\end{align*}
Recall that
\begin{align*}
    \mathcal{E}_{\varepsilon t}(v)=\Vert v \Vert_{\mathcal{Q}_W}^2 + \int_{\mathbb{R}^3} B_t \vert v\vert^2 + \frac{\kappa}{2} \Vert v\Vert_4^4,
\end{align*}
and that $E_{\varepsilon t} = \mathcal{E}_{\varepsilon t}(u(t))$. Hence,
\begin{align*}
    \Vert u(t) \Vert_{\mathcal{Q}_W}^2 
    &\leq E_{\varepsilon t} + \sup_{s\in [0,1]} \Vert r_s \Vert_\infty \\
    &\leq E_0 + \int_0^1 \Vert \partial_s r_s \Vert_\infty ds + \sup_{s\in [0,1]} \Vert r_s \Vert_\infty.
\end{align*}
This uniform bound on the $\mathcal{Q}_W$-norm of the solution implies that we can iterate the local existence proof to extend to a global solution. 

\paragraph{AI statement:} During the preparation of this work, the authors used OpenAI's ChatGPT and Anthropic's Claude to assist with developing and critically examining mathematical arguments, identifying relevant literature, and revising parts of the exposition. The authors independently checked the mathematical arguments and bibliographic references retained in the final manuscript and take full responsibility for its content.

\paragraph{Acknowledgements.}  SB acknowledges financial support from NSERC of Canada under Grant RGPIN-2024-03927. SS was funded by the Deutsche Forschungsgemeinschaft (DFG, German Research Foundation) under Germany’s Excellence Strategy – EXC 2111 – 390814868 (Munich Center for Quantum Science and Technology, MCQST) and the grant
TRR 352–Project-ID 470903074. 

\bibliography{biblio}

\begin{thebibliography}{10}

\bibitem{BornFock}
Max Born and Vladimir Fock.
\newblock Beweis des {Adiabatensatzes}.
\newblock {\em Zeitschrift f{\"u}r Physik}, 51(3--4):165--180, 1928.

\bibitem{kato1950adiabatic}
Tosio Kato.
\newblock On the adiabatic theorem of quantum mechanics.
\newblock {\em Journal of the Physical Society of Japan}, 5(6):435--439, 1950.

\bibitem{Avron2012}
Joseph~E. Avron, Martin Fraas, Gian-Michele Graf, and Philip Grech.
\newblock Adiabatic theorems for generators of contracting evolutions.
\newblock {\em Communications in Mathematical Physics}, 314(1):163--191, 2012.

\bibitem{bachmann2018adiabatic}
Sven Bachmann, Wojciech De~Roeck, and Martin Fraas.
\newblock The adiabatic theorem and linear response theory for extended quantum systems.
\newblock {\em Communications in Mathematical Physics}, 361:997--1027, 2018.

\bibitem{NS}
Bruno Nachtergaele and Robert Sims.
\newblock {Lieb-Robinson} bounds and the exponential clustering theorem.
\newblock {\em Communications in Mathematical Physics}, 265(1):119--130, 2006.

\bibitem{hastings2004locality}
Matthew~B. Hastings.
\newblock Locality in quantum and {Markov} dynamics on lattices and networks.
\newblock {\em Physical Review Letters}, 93:140402, Sep 2004.

\bibitem{BMNS}
Sven Bachmann, Spyridon Michalakis, Bruno Nachtergaele, and Robert Sims.
\newblock Automorphic equivalence within gapped phases of quantum lattice systems.
\newblock {\em Communications in Mathematical Physics}, 309(3):835--871, 2012.

\bibitem{MonacoTeufel}
Domenico Monaco and Stefan Teufel.
\newblock Adiabatic currents for interacting electrons on a lattice.
\newblock {\em Reviews in Mathematical Physics}, 31(03):1950009, 2019.

\bibitem{LemmRubiliani}
Marius Lemm and Carla Rubiliani.
\newblock {Lieb-Robinson bounds for Bose-Hubbard Hamiltonians: A review with a simplified proof}.
\newblock {\em arXiv preprint arXiv:2603.26209}, 2026.

\bibitem{lieb2002proof}
Elliott~H Lieb and Robert Seiringer.
\newblock Proof of {B}ose-{E}instein condensation for dilute trapped gases.
\newblock {\em Physical review letters}, 88(17):170409, 2002.

\bibitem{erdHos2010derivation}
L{\'a}szl{\'o} Erd{\H{o}}s, Benjamin Schlein, and Horng-Tzer Yau.
\newblock Derivation of the {G}ross-{P}itaevskii equation for the dynamics of {B}ose-{E}instein condensate.
\newblock {\em Annals of Mathematics}, pages 291--370, 2010.

\bibitem{ESYNoTrap}
L{\'a}szl{\'o} Erd{\H{o}}s, Benjamin Schlein, and Horng-Tzer Yau.
\newblock Derivation of the cubic non-linear schr{\"o}dinger equation from quantum dynamics of many-body systems.
\newblock {\em Inventiones mathematicae}, 167(3):515--614, 2006.

\bibitem{ErdosSchleinYau}
L{\'a}szl{\'o} Erd{\H{o}}s, Benjamin Schlein, and Horng-Tzer Yau.
\newblock Rigorous derivation of the gross-pitaevskii equation with a large interaction potential.
\newblock {\em Journal of the American Mathematical Society}, 22(4):1099--1156, 2009.

\bibitem{Pickl2010}
Peter Pickl.
\newblock Derivation of the time dependent gross--pitaevskii equation without positivity condition on the interaction.
\newblock {\em Journal of Statistical Physics}, 140(1):76--89, 2010.

\bibitem{carles2009nonlinear}
R{\'e}mi Carles.
\newblock Nonlinear {Schr\"odinger} equation with time dependent potential.
\newblock {\em Communications in Mathematical Sciences}, 9(4):937--964, 2011.

\bibitem{carles2015sharp}
R{\'e}mi Carles.
\newblock {Sharp weights in the Cauchy problem for nonlinear Schr{\"o}dinger equations with potential}.
\newblock {\em Zeitschrift f{\"u}r angewandte Mathematik und Physik}, 66(4):2087--2094, 2015.

\bibitem{GangGrech}
Zhou Gang and Philip Grech.
\newblock Adiabatic theorem for the {Gross–Pitaevskii} equation.
\newblock {\em Communications in Partial Differential Equations}, 42(5):731--756, 2017.

\bibitem{fermanian2020nonlinear}
Clotilde Fermanian-Kammerer and Alain Joye.
\newblock A nonlinear quantum adiabatic approximation.
\newblock {\em Nonlinearity}, 33(9):4715, 2020.

\bibitem{carles2011nonlinear}
R{\'e}mi Carles and Clotilde Fermanian-Kammerer.
\newblock A nonlinear adiabatic theorem for coherent states.
\newblock {\em Nonlinearity}, 24(8):2143, 2011.

\bibitem{hari2013coherent}
Lysianne Hari.
\newblock Coherent states for systems of l 2-supercritical nonlinear {S}chr{\"o}dinger equations.
\newblock {\em Communications in Partial Differential Equations}, 38(3):529--573, 2013.

\bibitem{sparber2016weakly}
Christof Sparber.
\newblock Weakly nonlinear time-adiabatic theory.
\newblock In {\em Annales Henri Poincar{\'e}}, volume~17, pages 913--936. Springer, 2016.

\bibitem{frohlich2004mean}
J{\"u}rg Fr{\"o}hlich and Enno Lenzmann.
\newblock Mean-field limit of quantum {B}ose gases and nonlinear {H}artree equation.
\newblock {\em S{\'e}minaire {\'E}quations aux d{\'e}riv{\'e}es partielles (Polytechnique) dit aussi" S{\'e}minaire Goulaouic-Schwartz"}, pages 1--26, 2004.

\bibitem{Temam1979}
Roger Temam.
\newblock {\em Navier--Stokes Equations: Theory and Numerical Analysis}, volume~2 of {\em Studies in Mathematics and Its Applications}.
\newblock North-Holland, Amsterdam, revised edition, 1979.

\bibitem{FarisSimon}
William~G. Faris and Barry Simon.
\newblock Degenerate and non-degenerate ground states for schr{\"o}dinger operators.
\newblock {\em Duke Mathematical Journal}, 42(3):589--618, 1975.

\bibitem{ReedSimonIV}
Michael Reed and Barry Simon.
\newblock {\em Methods of Modern Mathematical Physics: {IV}: {A}nalysis of Operators}.
\newblock Academic Press, New York, 1978.

\bibitem{YajimaZhang2004}
Kenji Yajima and Guoping Zhang.
\newblock Local smoothing property and {Strichartz} inequality for {Schr\"odinger} equations with potentials superquadratic at infinity.
\newblock {\em Journal of Differential Equations}, 202(1):81--110, 2004.

\bibitem{Mizutani2014}
Haruya Mizutani.
\newblock Strichartz estimates for {Schr{\"o}dinger} equations with variable coefficients and unbounded potentials {II}. superquadratic potentials.
\newblock {\em Communications on Pure and Applied Analysis}, 13:2177--2210, 2014.

\bibitem{simon1982schrodinger}
Barry Simon.
\newblock Schr{\"o}dinger semigroups.
\newblock {\em Bulletin of the American Mathematical Society}, 7(3):447--526, 1982.

\end{thebibliography}
\bibliographystyle{unsrt}

\end{document}